\documentclass[lettersize,journal]{IEEEtran}

\usepackage{etoolbox}
\makeatletter
\patchcmd{\@maketitle}{\Huge}{\fontsize{24}{34}\selectfont}{}{}
\makeatother

\usepackage{amsmath,amssymb,amsfonts}
\usepackage{mathtools}
\usepackage{array}

\usepackage{booktabs}
\usepackage{multirow}
\usepackage{bigdelim}
\usepackage{siunitx}

\usepackage{graphicx}
\usepackage{caption}
\usepackage{subcaption}
\usepackage[ruled,vlined,linesnumbered]{algorithm2e}
\SetAlgoVlined
\LinesNumbered
\SetKwInOut{KwIn}{Input}
\SetKwInOut{KwOut}{Output}
\SetKwComment{tcp}{// }{}
\SetKwFunction{Switch}{SwitchBus}
\SetKwFunction{Assemble}{AssembleResidual}
\SetKwFunction{EditJ}{EditJacobian}

\usepackage{microtype}
\usepackage{stfloats}
\usepackage{textcomp}
\usepackage{verbatim}
\usepackage{indentfirst}
\usepackage{enumitem}
\usepackage{balance}

\usepackage[dvipsnames]{xcolor}
\definecolor{ACgreen}{HTML}{2E7D32}
\definecolor{acblue}{HTML}{007FFF}
\definecolor{acorange}{HTML}{EB9800}
\definecolor{darkgreen}{rgb}{0.0,0.5,0.0}

\usepackage{tikz}
\usetikzlibrary{decorations.pathreplacing,calc}

\usepackage{cite}

\usepackage{url}
\usepackage[
    colorlinks=true,
    linkcolor=blue,
    citecolor=blue,
    urlcolor=black
]{hyperref}

\newcounter{model}
\renewcommand{\themodel}{\arabic{model}}
\newcommand{\modelline}[1]{%
  \refstepcounter{model}\label{#1}\text{\textbf{Model \themodel:}}%
}

\usepackage{calc}
\newlength{\forallw}
\def\BibTeX{{\rm B\kern-.05em{\sc i\kern-.025em b}\kern-.08em
    T\kern-.1667em\lower.7ex\hbox{E}\kern-.125emX}}

\begin{document}






\title{PACR: Parameter-Optimized AC Power Flow Restoration for AC Feasible DCOPF Dispatch}


\author{Michael A.~Boateng,~\IEEEmembership{Student Member,~IEEE,}
Russell Bent,~\IEEEmembership{Senior Member,~IEEE,}\\
Sidhant Misra,~\IEEEmembership{Senior Member,~IEEE,} Parikshit Pareek,~\IEEEmembership{Member,~IEEE,}\\
Pascal Van Hentenryck,~\IEEEmembership{Senior Member,~IEEE,} and Daniel K. Molzahn,~\IEEEmembership{Senior Member,~IEEE}



\thanks{}

}

\markboth{}
{Boateng \MakeLowercase{\textit{et al.}}: Restoring AC Feasibility of DCOPF Solutions via Parameter-Optimized Power Flow}

\maketitle

\begin{abstract}
The DC optimal power flow is widely used in power system operations because of its computational efficiency and scalability. However, DC dispatches are not guaranteed to satisfy the nonlinear AC power-flow equations or associated operational limits. This paper develops a parameterized, differentiable AC power-flow restoration method for mapping DC dispatches to AC-consistent operating points. The method incorporates distributed slack for active-power balancing and PV/PQ switching for reactive-power regulation, both implemented using smooth differentiable surrogates with tunable parameters, including slack participation factors, voltage setpoints, and regulation steepness. These parameters are trained offline by differentiating through the AC restoration equations using the implicit function theorem. Once trained, the optimized parameters are fixed and used directly during AC power-flow recovery from DC dispatches. The approach is evaluated on IEEE, ACTIVSg, and PEGASE test systems using setpoints computed by standard DC optimal power flow. Results show that the optimized restoration method improves AC feasibility recovery across various systems relative to conventional single-slack AC power-flow recovery. On the 9,241-bus case, the optimized method improves cost difference by $\mathbf{80\%}$ relative to the conventional recovery baseline and improves solving time relative to ACOPF by $\mathbf{75\%}$.
\end{abstract}


\vspace{-0.9em}

\section*{Nomenclature}

\subsection{Parameters:}

\begin{tabbing}
    $\mathbf{p}_{\mathrm{g}}^{\min},\, \mathbf{p}_{\mathrm{g}}^{\max}$ \hspace{1.5em} \= \kill
    ${\mathcal{N}}$ \> Set of buses; ${\mathcal{N}} = \{1,\, \dots,\, N\}$ \\
    $\mathcal{E}$ \> Set of lines $(i,j)$; reverse direction $(j,i)$ \\
    $\mathcal{G},\, \mathcal{L}$ \> Set of generators and loads; $\mathcal{G},\, \mathcal{L} \subseteq \mathcal{N}$ \\
    $\mathbf{r},\, \mathbf{b},\, \mathbf{g}$ \> Line resistance, susceptance, conductance \\ 
    $\mathbf{Y}$ \> Complex branch line admittance \\ 
    $\mathbf{\Phi}$ \> Line-Bus PTDF matrix, size $\mathcal{E} \times \mathcal{N}$ \\
    $\mathbf{s}_{\mathrm{tx}}^{\max}$ \> Apparent power transformer limit \\ 
    $\mathbf{i}_{\mathrm{line}}^{\max}$ \> Current flow line limit \\ 
    $\mathbf{p}_{\mathrm{g}}^{\min},\, \mathbf{p}_{\mathrm{g}}^{\max}$ \> Active power generation limits \\ 
    $\mathbf{q}_{\mathrm{g}}^{\min},\, \mathbf{q}_{\mathrm{g}}^{\max}$ \> Reactive power generation limits \\ 
    $\mathbf{v}^{\min},\, \mathbf{v}^{\max}$ \> Voltage magnitude limits \\
    $\epsilon$ \> Numerical convergence tolerance \\
\end{tabbing}

\vspace{-2.6em}

\subsection{Variables:}

\begin{tabbing}
    $\boldsymbol{\theta}^{\mathrm{dc}},\, \boldsymbol{\theta}^{\mathrm{ac}}$ \hspace{2.7em} \= \kill
    $\mathbf{p}_{\mathrm{g}},\, \mathbf{q}_{\mathrm{g}}$ \> Active and reactive power generation \\ 
    $\mathbf{s}_{\mathrm{g}}$ \> Complex power generation; $\mathbf{p}_{\mathrm{g}} + \mathbf{j} \cdot \mathbf{q}_{\mathrm{g}}$ \\ 
    $\mathbf{p}_{\mathrm{d}},\, \mathbf{q}_{\mathrm{d}}$ \> Active and reactive power demand \\ 
    $\boldsymbol{\theta}^{\mathrm{dc}},\, \boldsymbol{\theta}^{\mathrm{ac}}$ \> Voltage angles from DC and AC models \\ 
    $\ell$ \> Total active-power losses across network \\
\end{tabbing}

\vspace{-1.5em}

\section{Introduction}

\IEEEPARstart{T}{he} DC optimal power flow (DCOPF) problem is a core computational tool in power system operations, markets, and planning due to its tractability, scalability, and economic alignment with time-critical decision-making~\cite{stottdcpowerflow}.  The DCOPF is valuable in settings requiring repeated solves across large scenario sets, fast contingency screening, or integration within learning loops~\cite{chen2024endtoend,Ferrando2024PhysicsInformedTEMPR,Li2023HardLinearConstraints, chen2025optimallylinearizingpowerflow, TaheriMolzahn2024ACRestoration, taheri2024optimizing,constanteflores2024acnetwork, donti2021dc3, chen2021deepopf}. However, DCOPF solutions often violate the nonlinear AC power flow equations and/or operational constraints such as voltage limits and reactive power capabilities when these DCOPF setpoints are evaluated with an AC power flow (ACPF)~\cite{baker2021solutions,Fang2022,WangYang2026CoOptimizeDispatchPricing,stottdcpowerflow}.
To address this challenge, this paper develops a structured learning approach for \emph{AC feasibility restoration}.

\begin{figure}[!t]

\centering

\begin{minipage}{\linewidth}
\centering

\captionsetup{justification=centering,singlelinecheck=false}
\captionof{table}{\small\scshape Tunable parameters in the PACR Method}
\label{tab:trained_params}

\setlength{\tabcolsep}{4pt}
\renewcommand{\arraystretch}{1.1}
\resizebox{\columnwidth}{!}{%
\begin{tabular}{ll}
\toprule
\textbf{Parameter (role)} & \textbf{Interpretation} \\
\midrule
$\pi$ (softplus steepness) &
Smooth headroom map for slack weights ($\tilde h_i$) \\
$\phi$ (softmax temperature) &
Smooth participation-factor map ($\alpha_i$) from headroom \\
$\psi_s^{\dagger}$ (sigmoid steepness) &
Smooth PV/PQ regulation / PV--PQ surrogate \\
$\psi_t$ (tolerance band) &
Margin of reactive limits for stable switching \\
$ v_i^{\mathrm{sp}\dagger}$ (voltage setpoint) &
Tunable reference used in the PV/PQ surrogate \\
\bottomrule
\end{tabular}%
}

\vspace{0.35em}

\parbox{\columnwidth}{\footnotesize
$^{\dagger}$ The learned quantities are the voltage setpoints $\mathbf{v}^{\mathrm{sp}}$ and softmax temperature $\phi$ influencing participation factors. Steepness parameters $\{\pi,\psi_s,\psi_t\}$ are selected from empirically stable intervals for robust Newton convergence.}

\vspace{0.5em}
\end{minipage}

\vspace{0.8em}

\begin{minipage}{\linewidth}
\noindent\rule{\linewidth}{0.6pt}

\vspace{-0.2em}
\noindent \modelline{mod:dcopf}\ DC Optimal Power Flow (DCOPF)

\vspace{-0.6em}
\noindent\rule{\linewidth}{0.3pt}

\vspace{0.15em}

\vspace{-0.8em}

\begin{small}
\begin{subequations}
\label{eq:DCOPF}
\begin{align}
   & \hspace{-0.8em}\min \sum_{i \in \mathcal{G}} \big( c_{2,i}\,p_{\mathrm{g},i}^2 + c_{1,i}\,p_{\mathrm{g},i} + c_{0,i} \big)
&& \text{(Cost Minimization)} \label{eq:DCOPF-cost} \\[0em]
    & \hspace{-0.8em} \textbf{s.t. } p_{\mathrm{g},i} - p_{\mathrm{d},i} = \sum_{j \in \mathcal{E}} \overrightarrow{{p}_{j,i}} - \sum_{j \in \mathcal{E}} \overrightarrow{{p}_{i,j}}
    && \forall i \in \mathcal{N} \text{ (Balance Eqn.)} \label{eq:DCOPF-balance} \\[0em]
    & \overrightarrow{{p}_{i,j}} = b_{i,j}(\theta_i^{\mathrm{dc}} - \theta_j^{\mathrm{dc}})
    && \forall (i,j) \in \mathcal{E} \text{ (DC Flow)} \label{eq:DCOPF-flow} \\
    & |\overrightarrow{{p}_{i,j}}| \leq p_{i,j}^{\max}
    && \forall (i,j) \in \mathcal{E} \text{ (Line Limit)} \label{eq:DCOPF-plim} \\
    & \mathbf{p}_{\mathrm{g}}^{\min} \leq \mathbf{p}_{\mathrm{g}}^{\mathrm{dc}} \leq \mathbf{p}_{\mathrm{g}}^{\max}
    && \text{ (Gen. Active Power)} \label{eq:DCOPF-gen}
\end{align}
\end{subequations}
\end{small}

\vspace{-0.9em}
\noindent\rule{\linewidth}{0.3pt}
\end{minipage}

\label{fig:model1}

\vspace{1em}

\begin{minipage}{\linewidth}
\noindent\rule{\linewidth}{0.6pt}

\vspace{-0.2em}
\noindent \modelline{mod:acpf}\ Parameter-Optimized ACPF Restoration (PACR)

\vspace{-0.8em}
\noindent\rule{\linewidth}{0.3pt}

\vspace{0.15em}
\vspace{-1.2em}

\begin{small}
\begin{subequations}
\label{eq:ACPF}
\begin{align}
    & \hspace{-0.8em}\textbf{Power Injection Mismatches:} \nonumber \\
    & \Delta p_i =
    \underbrace{p_{\mathrm{g},i}^{\mathrm{dc}} + \alpha_i \cdot \ell}_{\mathclap{\text{smooth distributed slack}}}
    - p_{\mathrm{d},i} \nonumber \\
    & \qquad\qquad - \sum_{j \in \mathcal{N}} v_i v_j (g_{i,j}\cos\theta_{i,j} + b_{i,j}\sin\theta_{i,j})
    && \forall i \in \mathcal{N} \label{eq:ACPF-Pmis} \\
    & \Delta q_i = \underbrace{q_{\mathrm{g},i}^{\mathrm{ac}}(v_i)}_{\mathclap{\text{smooth PV/PQ}}} - q_{\mathrm{d},i} \nonumber \\
    & \qquad\qquad - \sum_{j \in \mathcal{N}} v_i v_j (g_{i,j}\sin\theta_{i,j} - b_{i,j}\cos\theta_{i,j})
    && \forall i \in \mathcal{N} \label{eq:ACPF-Qmis} \\[0.3em]
    & \hspace{-0.8em}\textbf{Smooth Distributed Slack (Participation)$^{*}$:} \nonumber \\
    & \tilde{h}_i = \frac{1}{\pi}\ln(1 + e^{\pi (p_{\mathrm{g},i}^{\max} - p_{\mathrm{g},i}^{\mathrm{dc}})}) - \frac{\ln 2}{\pi}
    && \forall i \in \mathcal{G} \label{eq:ACPF-headroom} \\
    & \alpha_i = \frac{e^{\phi \tilde{h}_i}}{\sum_{j \in \mathcal{G}} e^{\phi \tilde{h}_j}}, \quad \sum_{i \in \mathcal{G}} \alpha_i = 1
    && \forall i \in \mathcal{G} \label{eq:ACPF-participation} \\[0.3em]
& \hspace{-0.8em}\textbf{Smooth PV/PQ Control$^{*}$:} \nonumber \\
& \underbrace{
q_{\mathrm{g},i}^{\max} \gets q_{\mathrm{g},i}^{\max} - \psi_t,\quad
q_{\mathrm{g},i}^{\min} \gets q_{\mathrm{g},i}^{\min} + \psi_t
}_{\mathclap{\text{updated reactive-power limits}}}
\nonumber \\
& q_{\mathrm{g},i}^{\mathrm{ac}} = q_{\mathrm{g},i}^{\min} + \frac{q_{\mathrm{g},i}^{\max} - q_{\mathrm{g},i}^{\min}}{1 + e^{\psi_s(v_i - v_i^{\mathrm{sp}}) + \ln\left(\frac{q_{\mathrm{g},i}^{\max} - q_{\mathrm{g},i}^{{\mathrm{sp}}}}{q_{\mathrm{g},i}^{{\mathrm{sp}}} - q_{\mathrm{g},i}^{\min}}\right)}}
&& \forall i \in \mathcal{G} \label{eq:ACPF-Qsmooth} \\[0.3em]
& \hspace{-0.8em}\textbf{Augmented Jacobian System:} \nonumber \\
& 
\underbrace{
\begin{bmatrix}
\frac{\partial \Delta \mathbf{p}}{\partial \boldsymbol{\theta}} 
& \frac{\partial \Delta \mathbf{p}}{\partial \mathbf{v}} 
& \boldsymbol{\alpha} \\
\frac{\partial \Delta \mathbf{q}}{\partial \boldsymbol{\theta}} 
& \frac{\partial \Delta \mathbf{q}}{\partial \mathbf{v}} 
& \mathbf{0} \\
\mathbf{0}^T & \mathbf{0}^T & 0
\end{bmatrix}
}_{\mathclap{\mathbf{J}_{\mathrm{ac}}}}
\begin{bmatrix}
\Delta \boldsymbol{\theta} \\[0.2em] 
\Delta \mathbf{v} \\ 
\Delta \ell
\end{bmatrix}
=
-
\begin{bmatrix}
\Delta \mathbf{p} \\[0.2em] 
\Delta \mathbf{q} \\ 
\ell
\end{bmatrix}
\label{eq:ACPF-jacobian}
\end{align}
\end{subequations}
\end{small}

\vspace{-0.5em}

\begin{small}
\textbf{Convergence: } $\|\Delta \mathbf{p}\|_\infty, \|\Delta \mathbf{q}\|_\infty < \boldsymbol{\epsilon}$.
\end{small}

\vspace{-0.4em}
\noindent\rule{\linewidth}{0.3pt}

\vspace{0.8em}

\scriptsize{$^{*}$ PACR by default uses smooth surrogates for distributed slack and PV/PQ control, and approaches the discrete logic of these controls \cite{BoatengBentMisraPareekVanHentenryckMolzahn2025TowardsACFeasibility}, as smoothing is removed; see Fig.~\ref{fig:three-pdf-across}.}

\end{minipage}
\label{fig:model2}

\vspace{-0.3em}

\end{figure}

In standard ACPF restoration of DCOPF dispatches, convergence and solve-time performance are strongly influenced by how active- and reactive-power infeasibilities are handled. Active-power imbalance is often resolved using a single-slack-bus formulation, but such formulations can struggle to recover AC feasibility from DCOPF approximations, often leading to constraint violations or infeasible restored states~\cite{BoatengBentMisraPareekVanHentenryckMolzahn2025TowardsACFeasibility, baker2021solutions, Fang2022}. On the reactive-power side, iterative solvers typically rely on discrete PV/PQ switching, which can incur excessive outer-loop recalculations when generators reach reactive limits, thereby increasing computational cost and creating numerical convergence difficulties~\cite{zhao2008bus, zeng2023pvpq}. 

Recent work has shown that both distributed slack and improved reactive-power adjustments can substantially strengthen AC restoration. In particular, our earlier work introduced a structured DCOPF$\rightarrow$ACPF restoration pipeline that incorporates distributed slack and PV/PQ switching directly within the AC restoration step, and demonstrated that doing so materially improves the recovery of AC-feasible operating points from DC solutions~\cite{BoatengBentMisraPareekVanHentenryckMolzahn2025TowardsACFeasibility}. More broadly, distributed-slack formulations spread active-power imbalance across multiple generators, helping to avoid individual generators from exceeding their limits during restoration~\cite{Bharatwaj2012IterativeDCOPF, Dhople2020DistributedSlack}, while homotopy methods and smoothing of reactive-power curves can reduce Newton-Raphson iterations by mitigating the repeated recalculations caused by discrete PV/PQ switching~\cite{jereminov2020feasibility, zeng2023pvpq, McNamara2022TwoStageHomotopy, agarwal2018continuous}. 

This paper builds directly on these structured restoration frameworks. Rather than introducing new restoration policies, it studies how the distributed-slack and PV/PQ policies can be parameterized, smoothed, and learned to yield an even more effective approach. This is important because the quality of the restoration outcome depends critically on parameter choices such as participation factors and voltage setpoints, and poor choices can lead to slow convergence or infeasible solutions~\cite{TaheriMolzahn2024ACRestoration, zamzam2020learning, van2025bus}. To address this challenge, this paper develops a supervised learning framework that optimizes restoration parameters offline using ACOPF solutions as ground truth. Smoothing is introduced to make the restoration method differentiable and the parameter-learning more tractable. In summary, the paper's main contributions are: 

\textbf{1)} Developing a parameter-optimized ACPF restoration (PACR) formulation with smooth approximations for distributed slack (softplus headroom, softmax participation factors) and reactive power regulation (sigmoid PV/PQ curves), replacing discrete logic used in standard formulations.

\textbf{2)} Designing an implicit-function theorem-based parameter learning approach for ACPF restoration, enabling gradient-based optimization of parameters for differentiable correction of DCOPF dispatches to AC feasible points.

\textbf{3)} Demonstrating the transferability of optimized restoration parameters from the smooth differentiable ACPF formulation to the discrete ACPF recovery framework, while showing that the smooth formulation provides the strongest overall trade-off between AC feasibility, dispatch accuracy, and computational efficiency relative to the prior discrete framework~\cite{BoatengBentMisraPareekVanHentenryckMolzahn2025TowardsACFeasibility}.

\indent{The remainder of the paper is organized as follows. Section~\ref{sec:parameterizing_acpf} reveals the parameter-optimized AC restoration (PACR) method. Section~\ref{sec:restoration_dcopf} describes the $\text{DCOPF}\!\rightarrow\!\text{ACPF}$ feasibility restoration pipeline. Section~\ref{sec:num} reports the numerical results, and Section~\ref{sec:conclusion} concludes and provides future research directions.}

\section{Parameterizing the ACPF Restoration method}
\label{sec:parameterizing_acpf}

\begin{figure*}[!t]
\centering
\includegraphics[width=\textwidth]{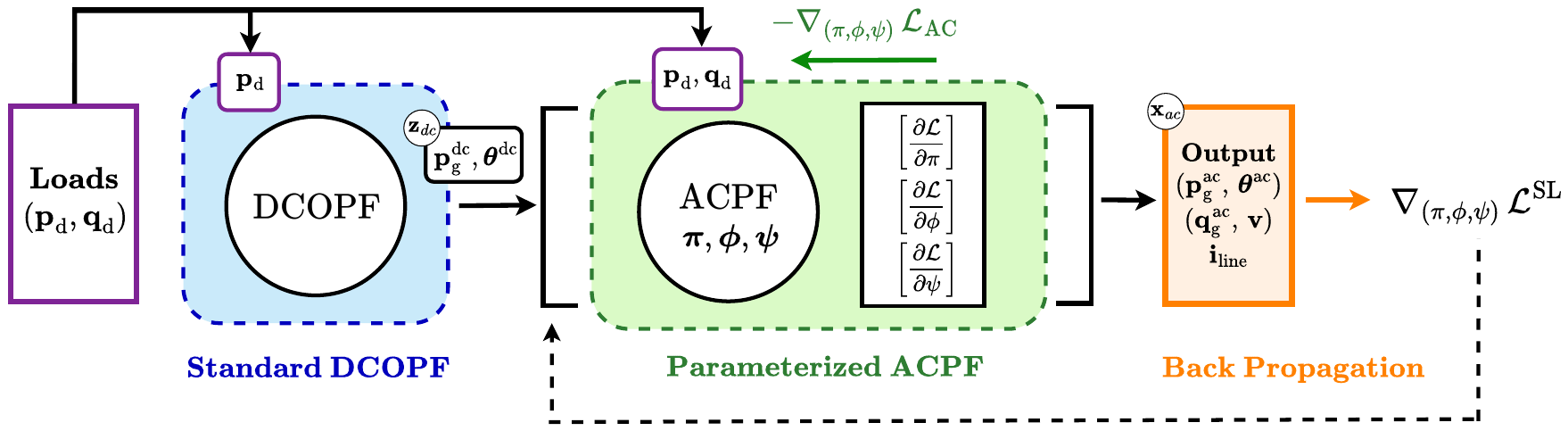}
\vspace{-1em}
\caption{\textcolor{violet}{\textbf{Load}} inputs $(\mathbf{p}_{\mathrm{d}},\mathbf{q}_{\mathrm{d}})$ feed a \textcolor{blue}{\textbf{DCOPF}} ($\scriptstyle \mathrm{DC}^{\textsc{BASE}}$) via $\mathbf{p}_{\mathrm{d}}$, producing DC setpoints $\textbf{z}_{dc}=(\mathbf{p}_{\mathrm{g}}^{\mathrm{dc}},\boldsymbol{\theta}^{\mathrm{dc}})$. These initialize the \textcolor{ACgreen}{\textbf{ACPF}} that maps $(\mathbf{p}_{\mathrm{d}},\mathbf{q}_{\mathrm{d}},\textbf{z}_{dc})\mapsto \textbf{x}_{ac}=\{(\mathbf{p}_{\mathrm{g}}^{\mathrm{ac}},\boldsymbol{\theta}^{\mathrm{ac}}),(\mathbf{q}_{\mathrm{g}}^{\mathrm{ac}},\mathbf{v}),\mathbf{i}_{\mathrm{line}}\}$. The ACPF contains switchable \emph{regulation controls}: \emph{1) Distributed slack} control with parameters $\boldsymbol{\pi}$ (softplus headroom steepness) and $\boldsymbol{\phi}$ (softmax participation temperature), and \emph{2) PV/PQ regulation} control with $\boldsymbol{\psi}=\{\boldsymbol{\psi}_s,\boldsymbol{\psi}_t,\mathbf{v}^{\mathrm{sp}}\}$ (sigmoid steepness, tolerance, and voltage setpoint). Both \emph{controls 1) \& 2)} are \emph{smooth} (S) implementations, thus yielding {\scriptsize $\mathrm{DCAC}^{\textsc{SS}}$}, and replacing the discrete pipelines {\scriptsize $\mathrm{DCAC}^{\textsc{DD}}$} \cite{BoatengBentMisraPareekVanHentenryckMolzahn2025TowardsACFeasibility}. \textcolor{orange}{\textbf{Outputs}} are trained with supervised learning (SL), using ACOPF targets.}
\label{fig:pipeline1}
  \vspace{-0.2em}
\end{figure*}

\begin{figure*}[!t]
  \centering
  \begin{subfigure}[t]{0.32\textwidth}
    \centering
    \includegraphics[width=\linewidth]{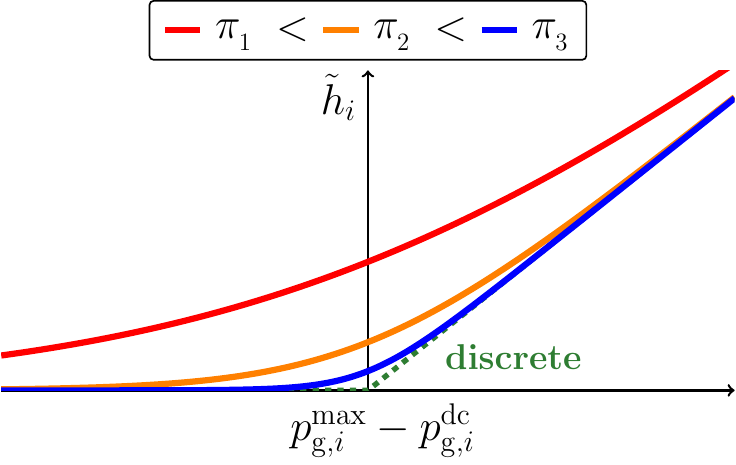}
    \caption{\text{Discrete} vs.\ \text{softplus} headroom ($\tilde{h}_i$)}
    \label{fig:pdf-a}
  \end{subfigure}\hfill
  \begin{subfigure}[t]{0.32\textwidth}
    \centering
    \includegraphics[width=\linewidth]{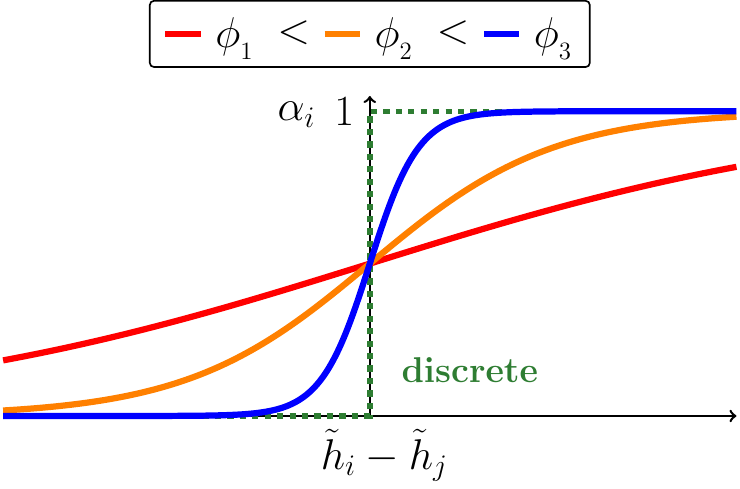}
    \caption{\text{Discrete} vs.\ \text{softmax} participation factor ($\alpha_i$)}
    \label{fig:pdf-b}
  \end{subfigure}\hfill
  \begin{subfigure}[t]{0.32\textwidth}
    \centering
    \includegraphics[width=\linewidth]{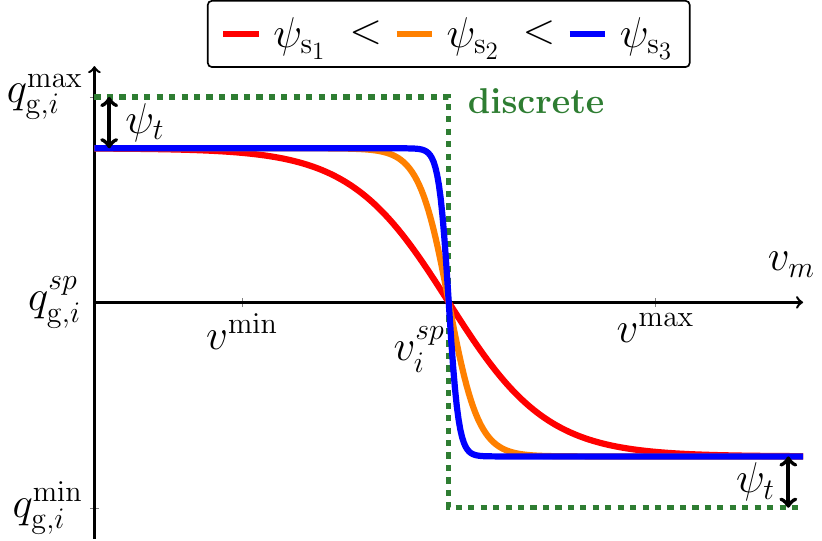}
    \caption{\text{Discrete} vs.\ \text{sigmoid} PV/PQ control curve}
    \label{fig:pdf-c}
  \end{subfigure}
  \caption{\text{Discrete vs.\ Smooth AC surrogates.}
    (a) Softplus headroom for slack allocation,
    (b) softmax participation factors, and
    (c) differentiable sigmoid PV/PQ control.
    \textcolor{ACgreen}{\textbf{Green}} denotes discrete baselines \cite{BoatengBentMisraPareekVanHentenryckMolzahn2025TowardsACFeasibility} and
    \textcolor{red}{\textbf{Red}}/\textcolor{orange}{\textbf{Orange}}/\textcolor{blue}{\textbf{Blue}}
    denote increasing smoothness. The smoothness adjustments are tuned via $(\pi,\phi,\psi_s)$ and shaped by $v^{\mathrm{sp}}$ (curve shift) and $\psi_t$ (Q-limit tolerance before $q_{\mathrm{g},i}^{\min}$/$q_{\mathrm{g},i}^{\max}$ limits engage).}
\label{fig:three-pdf-across}
  \vspace{-0.6em}
\end{figure*}

This section introduces the parameterized DCOPF$\rightarrow$ACPF restoration pipeline used in PACR. As shown in Fig.~\ref{fig:pipeline1}, load inputs first define the DCOPF problem, whose solution provides active-power dispatch and angle initializations for AC restoration. A parameterized ACPF model then restores the DC operating point to an AC-feasible state through two internal regulation modules: distributed slack for active-power balancing and PV/PQ regulation for reactive-power control. The key idea is to parameterize these internal restoration policies, so that the correction step itself becomes more effective and differentiable. Figure~\ref{fig:three-pdf-across} illustrates the corresponding smooth surrogates, and Table~\ref{tab:trained_params} summarizes the PACR parameters used.

We emphasize that all learned parameters are internal to the PACR method and are trained offline to improve the DCOPF$\rightarrow$ACPF correction. They do not modify operational AGC participation factors, AVR setpoints, or other controls in the actual system, nor do they model dynamic system response. The discussion below follows the two main restoration modules in turn: smooth distributed slack share for active-power balancing, and smooth PV/PQ switching for reactive-power and voltage control.

\vspace{-1em}

\subsection{Distributed Slack for Active Power Balancing}
\label{sec:distributed_slack}

The first lever in PACR is the distributed-slack mechanism embedded in Model~\ref{mod:acpf}, which is initialized from the standard DCOPF setpoints obtained from Model~\ref{mod:dcopf}. As also illustrated by the distributed-slack controls in the caption of Fig.~\ref{fig:pipeline1}, its role is to absorb the active-power imbalance that appears when the DCOPF dispatch is mapped into the nonlinear AC equations. Rather than assigning all imbalance to a single slack bus, the restoration distributes this correction across generators according to their available headroom \cite{BoatengBentMisraPareekVanHentenryckMolzahn2025TowardsACFeasibility,Bharatwaj2012IterativeDCOPF,Dhople2020DistributedSlack}. Prior work shows that headroom-aware slack allocation is effective at reducing post-restoration active-power violations \cite{BoatengBentMisraPareekVanHentenryckMolzahn2025TowardsACFeasibility}.

In the discrete setting, headroom and participation choices are assigned through non-smooth logic. Model~\ref{mod:acpf} replaces this with a differentiable two-stage mapping. First, the softplus surrogate in Fig.~\ref{fig:pdf-a} computes smooth generator headroom $\tilde{h}_i$ through~(\ref{eq:ACPF-headroom}). Second, the softmax mapping in Fig.~\ref{fig:pdf-b} converts these values into normalized participation factors $\alpha_i$ through~(\ref{eq:ACPF-participation}). The parameters $\pi$ and $\phi$ control the sharpness of these approximations and determine how closely the smooth model approaches the corresponding discrete rule. These participation factors enter directly into the active-power mismatch equations~(\ref{eq:ACPF-Pmis}) and the augmented Jacobian system~(\ref{eq:ACPF-jacobian}), allowing slack allocation and AC restoration to be solved simultaneously.

\vspace{-0.3em}

\subsection{PV/PQ Switching for Reactive Power and Voltage Control}
\label{sec:pv_pq_switching}

The second lever in PACR is the reactive-power regulation mechanism in Model~\ref{mod:acpf}, again applied after the standard DCOPF initialization from Model~\ref{mod:dcopf}. This part of the restoration replaces the conventional discrete PV/PQ switching rule used in previous AC power flow pipelines \cite{BoatengBentMisraPareekVanHentenryckMolzahn2025TowardsACFeasibility}. In the standard formulation, a generator remains in PV mode while it can regulate voltage within its reactive limits, and switches to PQ mode once one of those limits is reached. Although physically meaningful, this rule introduces discontinuities into the restoration map, making the ACPF harder to solve and unsuitable for gradient-based training \cite{jereminov2020feasibility,zeng2023pvpq,McNamara2022TwoStageHomotopy}. PACR replaces this discrete switching rule with a smooth surrogate.

Figure~\ref{fig:pdf-c} shows the transition from discrete PV/PQ switching to a differentiable sigmoid-based PV/PQ control curve. Instead of abruptly changing modes at $q_{\mathrm{g},i}^{\min}$ or $q_{\mathrm{g},i}^{\max}$, the smooth regulation in Model~\ref{mod:acpf}, given by~(\ref{eq:ACPF-Qsmooth}), creates a continuous relationship between voltage deviation and reactive-power response. The steepness parameter $\psi_s$ controls how sharply the sigmoid approximates the discrete switch, $\psi_t$ introduces a tolerance band near the reactive limits to improve numerical robustness, and the voltage setpoint $v_i^{\mathrm{sp}}$ shifts the curve to define the desired operating point. The logarithmic offset in~(\ref{eq:ACPF-Qsmooth}) ensures that when $v_i = v_i^{\mathrm{sp}}$, the reactive injection equals the desired setpoint $q_{\mathrm{g},i}^{\mathrm{sp}}$, as shown in Appendix~\ref{sec:appendix_a}. Embedding this smooth PV/PQ relation into the reactive mismatch equations~(\ref{eq:ACPF-Qmis}) and the augmented Jacobian system~(\ref{eq:ACPF-jacobian}) yields a tunable restoration procedure compatible with Newton-based AC correction.

\section{AC Restoration Learning Algorithm}
\label{sec:restoration_dcopf}



\begin{figure}[!t]
\centering
\includegraphics[width=0.49\textwidth]{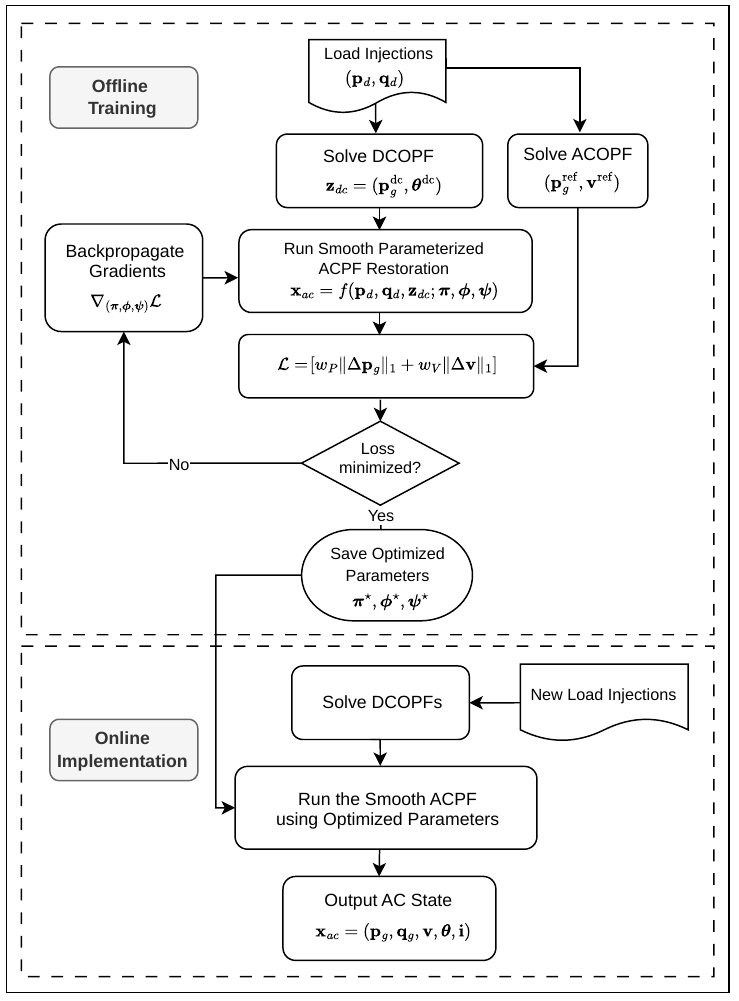}
\vspace{-0.9em}
\caption{Flowchart of the algorithm, illustrating the offline training and online implementation of the DCOPF$\rightarrow$ACPF pipeline.}
\label{fig:pipeline}
  \vspace{0.3em}
\end{figure}

The discussion below focuses on the optimization of the DCOPF$\rightarrow$ACPF restoration pipeline. In particular, we show how the parameter-optimized ACPF restoration (PACR) model can be trained to improve the differentiable correction of DCOPF dispatches to AC-feasible operating points through smooth distributed slack and smooth reactive-power regulation. Fig.~\ref{fig:pipeline} summarizes the offline--online workflow, and Algorithm~\ref{alg:acpf_train} outlines the corresponding parameter-optimization training procedure. The two main ingredients here are the sensitivity analysis required to differentiate through the converged augmented ACPF equations, and the supervised objective used to train PACR parameters against ACOPF reference solutions.

\subsection{Sensitivities of the DCOPF $\rightarrow$ ACPF Pipeline}
\label{sec:sensitivities_acpf}

The smooth PACR formulation enables gradient-based parameter optimization through the implicit function theorem (IFT). Starting from a DCOPF initialization $\mathbf{z}_{\mathrm{dc}}$ and reactive demand $\mathbf{q}_{\mathrm{d}}$, the augmented ACPF system is solved with Newton--Raphson using smooth distributed slack and smooth PV/PQ switching. At convergence, the restored AC state:
\[
\mathbf{x}_{\mathrm{ac}} = (\mathbf{p}_{\mathrm{g}}^{\mathrm{ac}}, \boldsymbol{\theta}^{\mathrm{ac}}, \mathbf{q}_{\mathrm{g}}^{\mathrm{ac}}, \mathbf{v})
\]
satisfies the following:
\[
\mathbf{g}(\mathbf{x}_{\mathrm{ac}}; \boldsymbol{\pi}, \boldsymbol{\phi}, \boldsymbol{\psi}) = \mathbf{0},
\]
where $\boldsymbol{\pi}$ governs the smooth headroom map, $\boldsymbol{\phi}$ controls participation factors, and $\boldsymbol{\psi} = \{\boldsymbol{\psi}_s, \boldsymbol{\psi}_t, \mathbf{v}^{\mathrm{sp}}\}$ defines the smooth PV/PQ regulation, and includes voltage setpoints.

Because the converged solution is defined implicitly by the augmented ACPF equations, its parameter sensitivities can be computed from the converged Jacobian $\mathbf{J}_{\mathrm{ac}}$ in~(\ref{eq:ACPF-jacobian}). For any parameter block $\boldsymbol{\xi} \in \{\boldsymbol{\pi}, \boldsymbol{\phi}, \boldsymbol{\psi}\}$,
\begin{equation}
\frac{\partial \mathbf{x}_{\mathrm{ac}}}{\partial \boldsymbol{\xi}} = -\mathbf{J}_{\mathrm{ac}}^{-1} \frac{\partial \mathbf{g}}{\partial \boldsymbol{\xi}}.
\label{eq:sensitivity}
\end{equation}
This reuses the same augmented Jacobian needed in the forward solve and avoids explicit unrolling of all Newton iterations. Additional derivations are given in Appendix~\ref{sec:appendix_b}.

\vspace{0.6em}

\subsection{Optimizing the Parameterized ACPF Restoration Method}
\label{sec:acpf_tuneable_params}

Parameter optimization requires careful loss function selection to balance AC feasibility, DC consistency, and computational efficiency. The $\mathrm{DCAC}^{\mathrm{SS}}$ formulation serves as the differentiable pipeline for computing optimal parameters $\boldsymbol{\pi}^\star, \boldsymbol{\phi}^\star, \boldsymbol{\psi}^\star$ through Algorithm~\ref{alg:acpf_train}. Supervised learning is used for training, with AC references $(\mathbf{p}_{\mathrm{g}}^{\mathrm{ref}}, \mathbf{v}^{\mathrm{ref}})$ obtained from ACOPF solutions. The full-batch loss over $\mathcal{D}$ is:
\vspace{-1em}

{\small
\begin{equation}
\mathcal{L}
=
\frac{1}{|\mathcal{D}|}
\sum_{\mathcal{D}}
\left[
w_P
\big\|\mathbf{p}_{\mathrm{g}}^{\mathrm{ac}}-\mathbf{p}_{\mathrm{g}}^{\mathrm{ref}}\big\|_1
+
w_V
\big\|\mathbf{v}-\mathbf{v}^{\mathrm{ref}}\big\|_1
\right].
\label{eq:loss_sl}
\end{equation}
}

The gradient-computation step in Algorithm~\ref{alg:acpf_train} combines the loss derivatives with the sensitivity analysis from~(\ref{eq:sensitivity}), enabling efficient parameter updates through Adam. The loss function~(\ref{eq:loss_sl}) is augmented with a reference regularization term
$w_{\mathrm{reg}}\!\left(\|\boldsymbol{\psi}-\boldsymbol{\psi}_{\mathrm{ref}}\|_2^2
+\|\boldsymbol{\pi}-\boldsymbol{\pi}_{\mathrm{ref}}\|_2^2
+\|\boldsymbol{\phi}-\boldsymbol{\phi}_{\mathrm{ref}}\|_2^2\right)$
to prevent overfitting and improve parameter stability. This term discourages large deviations from nominal and stable smoothness values rather than driving the parameters toward zero.
For evaluation, $\mathrm{DCAC}^{\text{BASE}}$ denotes the conventional single-slack restoration baseline without regulation or parameter optimization, while this paper's $\mathrm{DCAC}^{\text{SS}}$ and the $\mathrm{DCAC}^{\text{DD}}$~\cite{BoatengBentMisraPareekVanHentenryckMolzahn2025TowardsACFeasibility} incorporate smooth and discrete distributed-slack/PV--PQ regulation, respectively. Each method is tested with either \textit{unoptimized parameters} (\textit{init}) or \textit{optimized parameters} (\textit{opt}), where the latter are learned through Algorithm~\ref{alg:acpf_train}.

\vspace{-0.2em}

\renewcommand{\thealgocf}{1}
\begin{algorithm}[!t]
\setlength{\algomargin}{0.1em}
\small
\caption{\mbox{Adam Parameter Training using $\text{PACR}$}\protect\footnotemark}
\label{alg:acpf_train}
\DontPrintSemicolon
\SetKwInput{KwIn}{Input}
\SetKwInput{KwOut}{Output}
\SetKwInput{KwCons}{Hyperparameters}

\KwIn{%
  Training set $\mathcal{D}=\{(\mathbf{p}_{\mathrm{d}},\mathbf{q}_{\mathrm{d}},
  \mathbf{p}_{\mathrm{g}}^{\mathrm{ref}},\mathbf{v}^{\mathrm{ref}})\}$\\
  Fixed DCOPF map $\mathrm{DC}(\cdot)$ producing
  $\mathbf{z}_{\mathrm{dc}}=(\mathbf{p}_{\mathrm{g}}^{\mathrm{dc}},\boldsymbol{\theta}^{\mathrm{dc}})$\\
  Initial parameters $\boldsymbol{\pi}_1=\{\pi\}$,
  $\boldsymbol{\phi}_1=\{\phi\}$,
  $\boldsymbol{\psi}_1=\{\psi_s,\psi_t,\mathbf{v}^{\mathrm{sp}}\}$\\
  Voltage limits $\mathbf{v}^{\min},\mathbf{v}^{\max}$,
  gen.\ limits $\mathbf{p}_{\mathrm{g}}^{\min},\mathbf{p}_{\mathrm{g}}^{\max}$
}
\KwOut{%
  Trained parameters $\boldsymbol{\pi}^{\star},\boldsymbol{\phi}^{\star},\boldsymbol{\psi}^{\star}$
}
\KwCons{%
  Max iterations $N_{\mathrm{iter}}$; tolerance $\epsilon$; Adam stepsize $\eta$;
  moments $\beta_1,\beta_2$; weights $w_P,w_V,w_{\mathrm{reg}}$
}

Initialize Adam moments
$\mathbf{m}_{\xi},\mathbf{s}_{\xi}\gets\mathbf{0}$
for $\xi\in\{\boldsymbol{\pi},\boldsymbol{\phi},\boldsymbol{\psi}\}$\;

\For{$j=1,\dots,N_{\mathrm{iter}}$}{
  Initialize full-batch tracking loss $\mathcal{L}\gets0$\;
  
  \tcp{1) Full-Batch Forward Pass}
  \ForEach{$(\mathbf{p}_{\mathrm{d}},\mathbf{q}_{\mathrm{d}},
  \mathbf{p}_{\mathrm{g}}^{\mathrm{ref}},\mathbf{v}^{\mathrm{ref}})\in\mathcal{D}$}{
    $(\mathbf{p}_{\mathrm{g}}^{\mathrm{dc}},\boldsymbol{\theta}^{\mathrm{dc}})
    \gets \mathrm{DCOPF}$\;
    
    \mbox{$(\mathbf{p}_{\mathrm{g}}^{\mathrm{ac}},
    \mathbf{q}_{\mathrm{g}}^{\mathrm{ac}},\mathbf{v},\mathbf{i})
    \gets
    \mathrm{ACPF}_{\boldsymbol{\pi}_j,\boldsymbol{\phi}_j,\boldsymbol{\psi}_j}
    (\mathbf{p}_{\mathrm{d}},\mathbf{q}_{\mathrm{d}},
    \mathbf{p}_{\mathrm{g}}^{\mathrm{dc}},\boldsymbol{\theta}^{\mathrm{dc}})$}\;
    
$\mathcal{L}^{\mathrm{sc}}\gets
w_P
\big\|\mathbf{p}_{\mathrm{g}}^{\mathrm{ac}}
-\mathbf{p}_{\mathrm{g}}^{\mathrm{ref}}\big\|_1
+
w_V
\big\|\mathbf{v}-\mathbf{v}^{\mathrm{ref}}\big\|_1$\;
    
$\mathcal{L}\gets\mathcal{L}+\mathcal{L}^{\mathrm{sc}}$ \tcp{Scenario loss}
  }
  
  \tcp{2) Objective and Gradient Compute}
  $\mathcal{J}\leftarrow
  \dfrac{1}{|\mathcal{D}|}\mathcal{L}
  +w_{\mathrm{reg}}\left(
  \|\boldsymbol{\pi}_j-\boldsymbol{\pi}_{\mathrm{ref}}\|_2^2
  +
  \|\boldsymbol{\phi}_j-\boldsymbol{\phi}_{\mathrm{ref}}\|_2^2
  +
  \|\boldsymbol{\psi}_j-\boldsymbol{\psi}_{\mathrm{ref}}\|_2^2
  \right)$\;
  
Compute sensitivities and full-batch gradients of $\mathcal{J}$:\;
  \quad
  $\mathbf{g}_{\boldsymbol{\pi}}=\nabla_{\boldsymbol{\pi}}\mathcal{J},\quad
  \mathbf{g}_{\boldsymbol{\phi}}=\nabla_{\boldsymbol{\phi}}\mathcal{J},\quad
  \mathbf{g}_{\boldsymbol{\psi}}=\nabla_{\boldsymbol{\psi}}\mathcal{J}$\;
  
  \If{$\max\{\|\mathbf{g}_{\boldsymbol{\pi}}\|_\infty,
  \|\mathbf{g}_{\boldsymbol{\phi}}\|_\infty,
  \|\mathbf{g}_{\boldsymbol{\psi}}\|_\infty\}<\epsilon$}{
    \textbf{break} \tcp{Convergence achieved}
  }
  
  \tcp{3) Adam Parameter Update}
  \ForEach{$\xi\in\{\boldsymbol{\pi},\boldsymbol{\phi},\boldsymbol{\psi}\}$}{
    $\mathbf{m}_{\xi}\gets
    \beta_1\mathbf{m}_{\xi}+(1-\beta_1)\mathbf{g}_{\xi}$\;
    
    $\mathbf{s}_{\xi}\gets
    \beta_2\mathbf{s}_{\xi}
    +(1-\beta_2)(\mathbf{g}_{\xi}\odot\mathbf{g}_{\xi})$\;
    
    $\widehat{\mathbf{m}}_{\xi}\gets
    \mathbf{m}_{\xi}/(1-\beta_1^j)$,\quad
    $\widehat{\mathbf{s}}_{\xi}\gets
    \mathbf{s}_{\xi}/(1-\beta_2^j)$\;
    
    $\xi_{j+1}\gets
    \xi_j-\eta
    \dfrac{\widehat{\mathbf{m}}_{\xi}}
    {\sqrt{\widehat{\mathbf{s}}_{\xi}}+\varepsilon_{\mathrm{adam}}}$\;
  }
}
\KwRet{$\boldsymbol{\pi}^{\star}=\boldsymbol{\pi}_j,\ 
\boldsymbol{\phi}^{\star}=\boldsymbol{\phi}_j,\ 
\boldsymbol{\psi}^{\star}=\boldsymbol{\psi}_j$}\;

\BlankLine
\text{The optimized PACR parameters define $\mathrm{DCAC}^{\mathrm{SS}_{\text{opt}}}$.}\;
\end{algorithm}

\section{Numerical Experiments and Discussions} \label{sec:num}

The DCOPF$\rightarrow$ACPF pipeline described in Section~\ref{sec:acpf_tuneable_params} is evaluated through offline parameter optimization using supervised learning. This section presents the computational setup and results. Experiments were conducted on an IEEE test case \{\texttt{ieee\_118}\}, ACTIVSg networks \{\texttt{ACTIVSg\_200}, \texttt{ACTIVSg\_500}\}, and PEGASE networks \{\texttt{pegase\_1354}, \texttt{pegase\_2869}, \texttt{pegase\_9241}\}. For each case, $\mathrm{5{,}000}$ perturbed loading scenarios were generated, yielding $\mathrm{30{,}000}$ total scenarios across the six systems; these were split into $\mathrm{4{,}000}$ training and $\mathrm{1{,}000}$ test samples per case. Computations were carried out on the \texttt{Darwin} high-performance computing system at Los Alamos National Laboratory, using single $\mathrm{24}$-core compute nodes equipped with $\mathrm{32}$~GB RAM. ACOPF and DCOPF were modeled using \texttt{PowerModels.jl}~\cite{coffrin2018powermodels} and solved using \texttt{Ipopt}~\cite{wachter2006implementation}, while the ACPF implementation was developed using \texttt{pandapower.py}~\cite{thurner2018pandapower}. ACOPF was used as labeled targets for supervised learning. Active power demands ($\mathbf{p}_{\mathrm{d}}$) were perturbed using bounded Gaussian multiplicative noise, where the scaling factors were sampled from $\mathcal{N}(1.0,(15\%)^2)$. Reactive power demands ($\mathbf{q}_{\mathrm{d}}$) were then recomputed from randomly sampled power factors in $[0.95,1.0]$. ACOPF scenarios that failed to converge were excluded from the training dataset. The convergence rate exceeded $90\%$ across all ACPF variants with $\mathrm{DCAC}^{\text{BASE}}$ as setpoints. For computational efficiency, only voltage setpoints $\mathbf{v}^{\mathrm{sp}}$ for PV/PQ regulation and the softmax temperature $\phi$ for distributed-slack participation were optimized. The remaining smooth-control parameters $\{\pi,\psi_s,\psi_t\}$ were fixed to reference smoothing values selected from preliminary sweeps to ensure robust Newton-Raphson convergence. These reference values define moderately smooth approximations of the discrete controls, rather than serving as optimization variables. For example, appropriate PV/PQ sigmoid steepness selection reduced ACPF iterations from $\mathrm{35}$ to $\mathrm{4}$ on the \texttt{pegase\_1354} case.


Parameter optimization was performed using Adam from \texttt{scipy.optimize.minimize} with box constraints:
$\phi \in [50,900]$ for the participation temperature and $\mathbf{v}^{\mathrm{sp}} \in [0.95,1.05]$ for voltage setpoints. The regularization term penalizes deviations from the stable reference smoothing values, rather than driving the parameters toward zero. This discourages unnecessarily steep or ill-conditioned smooth approximations while allowing the optimized parameters to improve AC feasibility restoration. Loss function weights were used to select primary importance of generation dispatch and voltage profiles. Gradients were computed using ACPF derivatives. To maximize computational efficiency during optimization, the ACPF solver leverages sparse Jacobian construction with \texttt{scipy.sparse} CSC matrices, executing fast linear system updates via \texttt{scipy.sparse.linalg.spsolve}. This interface utilizes \texttt{SuperLU}, for large, nonsymmetric sparse systems via LU decomposition. Optimized parameters were deployed in both $\mathrm{DCAC}^{{\text{SS}}_{\text{opt}}}$ and $\mathrm{DCAC}^{{\text{DD}}_{\text{opt}}}$ variants to verify transferability of the training method. These optimized pipelines were compared against \textit{initial parameter values:} voltage setpoints initialized to $\mathrm{1.0}$ p.u. and the initial participation factors corresponding to a single-slack allocation, with participation factor $\mathrm{1.0}$ assigned to the designated slack bus and $\mathrm{0.0}$ assigned to other buses.  Newton-Raphson convergence tolerances was set to $10^{-6}$ p.u. Mean absolute error (MAE) and cost difference (CD) are accuracies relative to ACOPF ($|\mathcal{G}|$ is the number of generators):

\vspace{0.4em}

\begin{equation} \label{eq:MAE}
\mathrm{MAE} = \frac{1}{|\mathcal{G}|} 
\left\lVert \mathbf{p}_\mathrm{g} 
- \mathbf{p}_\mathrm{g}^{\mathrm{ACOPF}} \right\rVert_{1}.
\end{equation}

\begin{equation}\label{eq:cost_error}
\mathrm{CD}
= \frac{\left| \mathrm{Cost}^{\mathrm{DCAC}} 
- \mathrm{Cost}^{\mathrm{ACOPF}} \right|}
{\mathrm{Cost}^{\mathrm{ACOPF}}}\cdot 100.
\end{equation}

\vspace{0.5em}

\begin{table*}[!t]
\centering
\captionsetup{justification=centering,singlelinecheck=false}
\caption{{\small\scshape Average Constraint Violation Rates and Maximum Violation Magnitudes for $\mathrm{DCAC}^{\text{BASE}}$, $\mathrm{DCAC}^{\text{SS}_{\text{init}}}$, and $\mathrm{DCAC}^{\text{SS}_{\text{opt}}}$ across $\mathrm{1{,}000}$ Load Samples (Using $\mathrm{DC}_{\text{BASE}}$ setpoints)}}
\label{tab:big_eval_ss_viol_dcbase}
\LARGE

\setlength{\tabcolsep}{30pt}
\renewcommand{\arraystretch}{0.8}
\setlength{\extrarowheight}{1.2pt}
\setlength{\aboverulesep}{2pt}
\setlength{\belowrulesep}{1.5pt}

\resizebox{\textwidth}{!}{
\scalebox{1}[0.95]{
\begin{tabular}{llcccccccc}
\toprule
\multirow{2}{*}{\textbf{Test case}} & \multirow{2}{*}{\textbf{Method}} &
\multicolumn{2}{c}{\textbf{Active Power (p.u.)}} &
\multicolumn{2}{c}{\textbf{Reactive Power (p.u.)}} &
\multicolumn{2}{c}{\textbf{Voltage Limit (p.u.)}} &
\multicolumn{2}{c}{\textbf{Line Limit (\%)}} \\
\cmidrule(lr){3-4}\cmidrule(lr){5-6}\cmidrule(lr){7-8}\cmidrule(lr){9-10}
& &
\% viol. & Max & \% viol. & Max & \% viol. & Max & \% viol. & Max \\
\midrule

\multirow{3}{*}{\texttt{ieee\_118}}
& $\mathrm{DCAC}^{\text{BASE}}$
& $\mathrm{1.61}$ & $\mathrm{0.01}$ & $\mathrm{5.56}$ & $\mathrm{2.93}$ & $\mathrm{0.05}$ & $\mathrm{0.01}$ & $\mathrm{0.00}$ & $\mathrm{0.20}$ \\
& ${\mathrm{DCAC}^{{\text{SS}}_{\text{init}}}}$
& $\mathrm{0.00}$ & $\mathrm{0.00}$ & $\mathrm{0.00}$ & $\mathrm{0.00}$ & $\mathrm{0.00}$ & $\mathrm{0.00}$ & $\mathrm{0.00}$ & $\mathrm{0.00}$ \\
& ${\mathrm{DCAC}^{{\text{SS}}_{\text{opt}}}}$
& $\mathbf{0.00}$ & $\mathbf{0.00}$ & $\mathbf{0.00}$ & $\mathbf{0.00}$ & $\mathbf{0.00}$ & $\mathbf{0.00}$ & $\mathbf{0.00}$ & $\mathbf{0.00}$ \\
\midrule

\multirow{3}{*}{\texttt{ACTIVSg\_200}}
& $\mathrm{DCAC}^{\text{BASE}}$
& $\mathrm{0.00}$ & $\mathrm{0.00}$ & $\mathrm{0.00}$ & $\mathrm{0.00}$ & $\mathrm{0.00}$ & $\mathrm{0.00}$ & $\mathrm{0.00}$ & $\mathrm{0.00}$ \\
& ${\mathrm{DCAC}^{{\text{SS}}_{\text{init}}}}$
& $\mathrm{0.00}$ & $\mathrm{0.00}$ & $\mathrm{0.00}$ & $\mathrm{0.00}$ & $\mathrm{0.00}$ & $\mathrm{0.00}$ & $\mathrm{0.00}$ & $\mathrm{0.00}$ \\
& ${\mathrm{DCAC}^{{\text{SS}}_{\text{opt}}}}$
& $\mathbf{0.00}$ & $\mathbf{0.00}$ & $\mathbf{0.00}$ & $\mathbf{0.00}$ & $\mathbf{0.00}$ & $\mathbf{0.00}$ & $\mathbf{0.00}$ & $\mathbf{0.00}$ \\
\midrule

\multirow{3}{*}{\texttt{ACTIVSg\_500}}
& $\mathrm{DCAC}^{\text{BASE}}$
& $\mathrm{0.00}$ & $\mathrm{0.00}$ & $\mathrm{8.03}$ & $\mathrm{0.06}$ & $\mathrm{0.00}$ & $\mathrm{0.00}$ & $\mathrm{0.00}$ & $\mathrm{0.00}$ \\
& ${\mathrm{DCAC}^{{\text{SS}}_{\text{init}}}}$
& $\mathrm{0.00}$ & $\mathrm{0.00}$ & $\mathrm{0.00}$ & $\mathrm{0.00}$ & $\mathrm{0.00}$ & $\mathrm{0.00}$ & $\mathrm{0.03}$ & $\mathrm{0.71}$ \\
& ${\mathrm{DCAC}^{{\text{SS}}_{\text{opt}}}}$
& $\mathbf{0.00}$ & $\mathbf{0.00}$ & $\mathbf{0.00}$ & $\mathbf{0.00}$ & $\mathbf{0.00}$ & $\mathbf{0.00}$ & $\mathbf{0.00}$ & $\mathbf{0.00}$ \\
\midrule

\multirow{3}{*}{\texttt{pegase\_1354}}
& $\mathrm{DCAC}^{\text{BASE}}$
& $\mathrm{0.46}$ & $\mathrm{0.02}$ & $\mathrm{18.46}$ & $\mathrm{18.28}$ & $\mathrm{0.00}$ & $\mathrm{0.00}$ & $\mathrm{0.00}$ & $\mathrm{0.00}$ \\
& ${\mathrm{DCAC}^{{\text{SS}}_{\text{init}}}}$
& $\mathrm{0.00}$ & $\mathrm{0.00}$ & $\mathrm{0.00}$ & $\mathrm{0.00}$ & $\mathrm{0.00}$ & $\mathrm{0.00}$ & $\mathrm{0.00}$ & $\mathrm{0.00}$ \\
& ${\mathrm{DCAC}^{{\text{SS}}_{\text{opt}}}}$
& $\mathbf{0.00}$ & $\mathbf{0.00}$ & $\mathbf{0.00}$ & $\mathbf{0.00}$ & $\mathbf{0.00}$ & $\mathbf{0.00}$ & $\mathbf{0.00}$ & $\mathbf{0.00}$ \\
\midrule

\multirow{3}{*}{\texttt{pegase\_2869}}
& $\mathrm{DCAC}^{\text{BASE}}$
& $\mathrm{0.51}$ & $\mathrm{2.63}$ & $\mathrm{15.69}$ & $\mathrm{13.73}$ & $\mathrm{0.00}$ & $\mathrm{0.00}$ & $\mathrm{0.00}$ & $\mathrm{0.00}$ \\
& ${\mathrm{DCAC}^{{\text{SS}}_{\text{init}}}}$
& $\mathrm{0.00}$ & $\mathrm{0.00}$ & $\mathrm{0.00}$ & $\mathrm{0.00}$ & $\mathrm{0.21}$ & $\mathrm{0.02}$ & $\mathrm{0.00}$ & $\mathrm{0.00}$ \\
& ${\mathrm{DCAC}^{{\text{SS}}_{\text{opt}}}}$
& $\mathbf{0.00}$ & $\mathbf{0.00}$ & $\mathbf{0.00}$ & $\mathbf{0.00}$ & $\mathbf{0.03}$ & $\mathbf{0.01}$ & $\mathbf{0.00}$ & $\mathbf{0.00}$ \\
\midrule

\multirow{3}{*}{\texttt{pegase\_9241}}
& $\mathrm{DCAC}^{\text{BASE}}$
& $\mathrm{2.31}$ & $\mathrm{14.66}$ & $\mathrm{14.74}$ & $\mathrm{10.71}$ & $\mathrm{17.23}$ & $\mathrm{0.07}$ & $\mathrm{2.31}$ & $\mathrm{44.73}$ \\
& ${\mathrm{DCAC}^{{\text{SS}}_{\text{init}}}}$
& $\mathrm{0.00}$ & $\mathrm{0.00}$ & $\mathrm{0.00}$ & $\mathrm{0.00}$ & $\mathrm{3.75}$ & $\mathrm{0.02}$ & $\mathrm{0.00}$ & $\mathrm{0.00}$ \\
& ${\mathrm{DCAC}^{{\text{SS}}_{\text{opt}}}}$
& $\mathbf{0.00}$ & $\mathbf{0.00}$ & $\mathbf{0.00}$ & $\mathbf{0.00}$ & $\mathbf{2.61}$ & $\mathbf{0.02}$ & $\mathbf{0.10}$ & $\mathbf{0.05}$ \\
\bottomrule
\end{tabular}}}
\end{table*}

\begin{table*}[!t]
\centering
\captionsetup{justification=centering,singlelinecheck=false}
\caption{{\small\scshape Performance Table for $\mathrm{DCAC}^{\text{BASE}}$, $\mathrm{DCAC}^{\text{SS}_{\text{init}}}$, and $\mathrm{DCAC}^{\text{SS}_{\text{opt}}}$ across $\mathrm{1{,}000}$ samples ($\mathrm{DC}_{\text{BASE}}$ setpoints)}}
\label{tab:big_eval_ss_perf_dcbase}
\LARGE

\setlength{\tabcolsep}{26pt}
\renewcommand{\arraystretch}{0.8}
\setlength{\extrarowheight}{1.2pt}
\setlength{\aboverulesep}{3pt}
\setlength{\belowrulesep}{2.5pt}

\resizebox{\textwidth}{!}{
\scalebox{1}[0.95]{
\begin{tabular}{llcccccccc}
\toprule
\multirow{2}{*}{\textbf{Test case}} & \multirow{2}{*}{\textbf{Method}} &
\multicolumn{2}{c}{\textbf{Cost Diff. (\%)}} &
\multicolumn{2}{c}{\textbf{MAE$^{\star}$ (p.u.)}} & 
\multicolumn{2}{c}{\textbf{Iter. Count}} &
\multicolumn{2}{c}{\textbf{Solve Time (s)}} \\
\cmidrule(lr){3-4}\cmidrule(lr){5-6}\cmidrule(lr){7-8}\cmidrule(lr){9-10}
& &
val. & \%Improv. &
val. & \%Improv. &
val. & \%Improv. &
val. & \%Improv. \\
\midrule
\midrule

\multirow{3}{*}{\texttt{ieee\_118}}
& $\mathrm{DCAC}^{\text{BASE}}$
& $1.488$ & $\mathbf{+33.8}$ & $0.025$ & $\mathbf{+8.0}$ & $\mathrm{4.0}$ & $\text{---}$ & $0.073$ & $\text{---}$ \\
& ${\mathrm{DCAC}^{{\text{SS}}_{\text{init}}}}$
& $0.993$ & $\mathbf{+0.8}$ & $0.023$ & $0.0$ & $5.0$ & $-8.0$ & $0.269$ & $\mathbf{+24.2}$ \\
& ${\mathrm{DCAC}^{{\text{SS}}_{\text{opt}}}}$
& $0.985$ & $\text{---}$ & $0.023$ & $\text{---}$ & $5.4$ & $\text{---}$ & $0.204$ & $\text{---}$ \\
\midrule

\multirow{3}{*}{\texttt{ACTIVSg\_200}}
& $\mathrm{DCAC}^{\text{BASE}}$
& $0.035$ & $\text{---}$ & $0.003$ & $0.0$ & $\mathrm{2.0}$ & $\text{---}$ & $0.065$ & $\text{---}$ \\
& ${\mathrm{DCAC}^{{\text{SS}}_{\text{init}}}}$
& $0.524$ & $-13.9$ & $0.003$ & $0.0$ & $4.0$ & $\mathbf{+25.0}$ & $0.258$ & $\mathbf{+46.5}$ \\
& ${\mathrm{DCAC}^{{\text{SS}}_{\text{opt}}}}$
& $0.597$ & $\text{---}$ & $0.003$ & $\text{---}$ & $3.0$ & $\text{---}$ & $0.138$ & $\text{---}$ \\
\midrule

\multirow{3}{*}{\texttt{ACTIVSg\_500}}
& $\mathrm{DCAC}^{\text{BASE}}$
& $1.383$ & $\mathbf{+70.1}$ & $0.010$ & $0.0$ & $\mathrm{3.0}$ & $\text{---}$ & $0.113$ & $\text{---}$ \\
& ${\mathrm{DCAC}^{{\text{SS}}_{\text{init}}}}$
& $2.196$ & $\mathbf{+81.2}$ & $0.010$ & $0.0$ & $3.4$ & $\mathbf{+5.9}$ & $0.266$ & $\mathbf{+15.4}$ \\
& ${\mathrm{DCAC}^{{\text{SS}}_{\text{opt}}}}$
& $0.413$ & $\text{---}$ & $0.010$ & $\text{---}$ & $3.2$ & $\text{---}$ & $0.225$ & $\text{---}$ \\
\midrule

\multirow{3}{*}{\texttt{pegase\_1354}}
& $\mathrm{DCAC}^{\text{BASE}}$
& $0.394$ & $\mathbf{+88.8}$ & $0.050$ & $\mathbf{+20.0}$ & $\mathrm{4.0}$ & $\text{---}$ & $0.361$ & $\text{---}$ \\
& ${\mathrm{DCAC}^{{\text{SS}}_{\text{init}}}}$
& $0.333$ & $\mathbf{+86.8}$ & $0.048$ & $\mathbf{+16.7}$ & $4.8$ & $\mathbf{+10.4}$ & $1.102$ & $\mathbf{+19.9}$ \\
& ${\mathrm{DCAC}^{{\text{SS}}_{\text{opt}}}}$
& $0.044$ & $\text{---}$ & $0.040$ & $\text{---}$ & $4.3$ & $\text{---}$ & $0.883$ & $\text{---}$ \\
\midrule

\multirow{3}{*}{\texttt{pegase\_2869}}
& $\mathrm{DCAC}^{\text{BASE}}$
& $0.458$ & $\mathbf{+95.4}$ & $0.042$ & $\mathbf{+28.6}$ & $\mathrm{4.0}$ & $\text{---}$ & $0.915$ & $\text{---}$ \\
& ${\mathrm{DCAC}^{{\text{SS}}_{\text{init}}}}$
& $0.387$ & $\mathbf{+94.6}$ & $0.041$ & $\mathbf{+26.8}$ & $5.3$ & $\mathbf{+17.0}$ & $2.324$ & $\mathbf{+19.8}$ \\
& ${\mathrm{DCAC}^{{\text{SS}}_{\text{opt}}}}$
& $0.021$ & $\text{---}$ & $0.030$ & $\text{---}$ & $4.4$ & $\text{---}$ & $1.865$ & $\text{---}$ \\
\midrule

\multirow{3}{*}{\texttt{pegase\_9241}}
& $\mathrm{DCAC}^{\text{BASE}}$
& $0.631$ & $\mathbf{+80.3}$ & $0.038$ & $\mathbf{+34.2}$ & $\mathrm{3.0}$ & $\text{---}$ & $3.389$ & $\text{---}$ \\
& ${\mathrm{DCAC}^{{\text{SS}}_{\text{init}}}}$
& $0.510$ & $\mathbf{+75.7}$ & $0.035$ & $\mathbf{+28.6}$ & $7.1$ & $\mathbf{+15.5}$ & $10.516$ & $\mathbf{+25.3}$ \\
& ${\mathrm{DCAC}^{{\text{SS}}_{\text{opt}}}}$
& $0.124$ & $\text{---}$ & $0.025$ & $\text{---}$ & $6.0$ & $\text{---}$ & $7.860$ & $\text{---}$ \\
\bottomrule
\end{tabular}}}
\vspace{0.6em}
\begin{flushleft}
\vspace{-0.8em}
\scriptsize{

The \% Improv. report values relative to $\mathrm{DCAC}^{\text{SS}_{\text{opt}}}$ for each metric. $\mathrm{DCAC}^{\text{SS}_{\text{opt}}}$ uses optimized values for smooth distributed slack and smooth PV/PQ switching. The initialized smooth pipeline, $\mathrm{DCAC}^{\text{SS}_{\text{init}}}$, uses initial parameter values: voltage setpoints initialized to $\mathrm{1.0}$~p.u., with participation factor $\mathrm{1.0}$ assigned to the slack bus and $\mathrm{0.0}$ to all other buses.
}
\end{flushleft}
\vspace{-0.6em}
\end{table*}

\begin{table}[!t]
\centering
\captionsetup{justification=centering,singlelinecheck=false}
\caption{\scshape Test loss values across $\mathrm{1{,}000}$ samples comparing $\mathrm{DCAC}^{\text{BASE}}$, $\mathrm{DCAC}^{\text{SS}_{\text{init}}}$ and $\mathrm{DCAC}^{\text{SS}_{\text{opt}}}$ (Using $\mathrm{DC}_{\text{BASE}}$ setpoints)}
\label{tab:dcac_ss_loss}
\scriptsize
\setlength{\tabcolsep}{10pt}
\renewcommand{\arraystretch}{1.2}
\begin{tabular}{l ccc}
\toprule
\textbf{Test Case} & \multicolumn{3}{c}{\textbf{Evaluation Loss} $\mathcal{L}_{\mathrm{}}$} \\
\cmidrule(lr){2-4}
& ${}^{\star}\mathrm{DCAC}^{\text{BASE}}$
& ${}^{\star}\mathrm{DCAC}^{\text{SS}_{\text{init}}}$ 
& $\mathrm{DCAC}^{\text{SS}_{\text{opt}}}$ \\
\midrule
\midrule
\texttt{ieee\_118}     & $\mathrm{0.212}$ & $\mathrm{0.029}$ & $\mathbf{0.026}$ \\
\texttt{ACTIVSg\_200}  & $\mathrm{0.113}$ & $\mathrm{0.004}$ & $\mathbf{0.002}$ \\
\texttt{ACTIVSg\_500}  & $\mathrm{0.122}$ & $\mathrm{0.015}$ & $\mathbf{0.002}$ \\
\midrule
\texttt{pegase\_1354}  & $\mathrm{0.131}$ & $\mathbf{0.001}$ & $\mathrm{0.002}$ \\
\texttt{pegase\_2869}  & $\mathrm{0.145}$ & $\mathrm{0.041}$ & $\mathbf{0.031}$ \\
\texttt{pegase\_9241}  & $\mathrm{0.254}$ & $\mathrm{0.042}$ & $\mathbf{0.025}$ \\
\bottomrule
\end{tabular}
\vspace{-0.6em}
\begin{flushleft}
\scriptsize{${}^{\star}\mathrm{DCAC}^{\text{BASE}}$ and ${}^{\star}\mathrm{DCAC}^{\text{SS}_{\text{init}}}$ denote unoptimized pipelines; their losses are evaluated using~\eqref{eq:loss_sl} upon ACPF convergence.}
\end{flushleft}
\vspace{-0.5em}
\end{table}

\begin{table*}[!t]
\centering
\captionsetup{justification=centering,singlelinecheck=false}
\caption{{\small\scshape Average Constraint Violation Rates and Maximum Violation Magnitudes for $\mathrm{DCAC}^{\text{BASE}}$, $\mathrm{DCAC}^{\text{DD}_{\text{init}}}$, and $\mathrm{DCAC}^{\text{DD}_{\text{opt}}}$ across $\mathrm{1{,}000}$ Load Samples (Using $\mathrm{DC}_{\text{BASE}}$ setpoints)}}
\label{tab:big_eval_dd_viol_dcbase}
\LARGE

\setlength{\tabcolsep}{30pt}
\renewcommand{\arraystretch}{0.8}
\setlength{\extrarowheight}{1.2pt}
\setlength{\aboverulesep}{2pt}
\setlength{\belowrulesep}{1.5pt}

\resizebox{\textwidth}{!}{
\scalebox{1}[0.95]{
\begin{tabular}{llcccccccc}
\toprule
\multirow{2}{*}{\textbf{Test case}} & \multirow{2}{*}{\textbf{Method}} &
\multicolumn{2}{c}{\textbf{Active Power (p.u.)}} &
\multicolumn{2}{c}{\textbf{Reactive Power (p.u.)}} &
\multicolumn{2}{c}{\textbf{Voltage Limit (p.u.)}} &
\multicolumn{2}{c}{\textbf{Line Limit (\%)}} \\
\cmidrule(lr){3-4}\cmidrule(lr){5-6}\cmidrule(lr){7-8}\cmidrule(lr){9-10}
& &
\% viol. & Max & \% viol. & Max & \% viol. & Max & \% viol. & Max \\
\midrule

\multirow{3}{*}{\texttt{ieee\_118}}
& $\mathrm{DCAC}^{\text{BASE}}$
& $\mathrm{1.61}$ & $\mathrm{0.01}$ & $\mathrm{5.56}$ & $\mathrm{2.93}$ & $\mathrm{0.05}$ & $\mathrm{0.01}$ & $\mathrm{0.00}$ & $\mathrm{0.20}$ \\
& ${\mathrm{DCAC}^{{\text{DD}}_{\text{init}}}}$
& $\mathrm{0.00}$ & $\mathrm{0.00}$ & $\mathrm{0.00}$ & $\mathrm{0.00}$ & $\mathrm{0.00}$ & $\mathrm{0.00}$ & $\mathrm{0.00}$ & $\mathrm{0.00}$ \\
& ${\mathrm{DCAC}^{{\text{DD}}_{\text{opt}}}}$
& $\mathbf{0.00}$ & $\mathbf{0.00}$ & $\mathbf{0.00}$ & $\mathbf{0.00}$ & $\mathbf{0.00}$ & $\mathbf{0.00}$ & $\mathbf{0.00}$ & $\mathbf{0.00}$ \\
\midrule

\multirow{3}{*}{\texttt{ACTIVSg\_200}}
& $\mathrm{DCAC}^{\text{BASE}}$
& $\mathrm{0.00}$ & $\mathrm{0.00}$ & $\mathrm{0.00}$ & $\mathrm{0.00}$ & $\mathrm{0.00}$ & $\mathrm{0.00}$ & $\mathrm{0.00}$ & $\mathrm{0.00}$ \\
& ${\mathrm{DCAC}^{{\text{DD}}_{\text{init}}}}$
& $\mathrm{0.00}$ & $\mathrm{0.00}$ & $\mathrm{0.00}$ & $\mathrm{0.00}$ & $\mathrm{0.07}$ & $\mathrm{0.03}$ & $\mathrm{0.00}$ & $\mathrm{0.00}$ \\
& ${\mathrm{DCAC}^{{\text{DD}}_{\text{opt}}}}$
& $\mathbf{0.00}$ & $\mathbf{0.00}$ & $\mathbf{0.00}$ & $\mathbf{0.00}$ & $\mathbf{0.00}$ & $\mathbf{0.00}$ & $\mathbf{0.00}$ & $\mathbf{0.00}$ \\
\midrule

\multirow{3}{*}{\texttt{ACTIVSg\_500}}
& $\mathrm{DCAC}^{\text{BASE}}$
& $\mathrm{0.00}$ & $\mathrm{0.00}$ & $\mathrm{8.03}$ & $\mathrm{0.06}$ & $\mathrm{0.00}$ & $\mathrm{0.00}$ & $\mathrm{0.00}$ & $\mathrm{0.00}$ \\
& ${\mathrm{DCAC}^{{\text{DD}}_{\text{init}}}}$
& $\mathrm{0.00}$ & $\mathrm{0.00}$ & $\mathrm{0.00}$ & $\mathrm{0.00}$ & $\mathrm{0.00}$ & $\mathrm{0.00}$ & $\mathrm{0.16}$ & $\mathrm{1.46}$ \\
& ${\mathrm{DCAC}^{{\text{DD}}_{\text{opt}}}}$
& $\mathbf{0.00}$ & $\mathbf{0.00}$ & $\mathbf{0.00}$ & $\mathbf{0.00}$ & $\mathbf{0.00}$ & $\mathbf{0.00}$ & $\mathbf{0.00}$ & $\mathbf{0.00}$ \\
\midrule

\multirow{3}{*}{\texttt{pegase\_1354}}
& $\mathrm{DCAC}^{\text{BASE}}$
& $\mathrm{0.46}$ & $\mathrm{0.02}$ & $\mathrm{18.46}$ & $\mathrm{18.28}$ & $\mathrm{0.00}$ & $\mathrm{0.00}$ & $\mathrm{0.00}$ & $\mathrm{0.00}$ \\
& ${\mathrm{DCAC}^{{\text{DD}}_{\text{init}}}}$
& $\mathrm{0.00}$ & $\mathrm{0.00}$ & $\mathrm{0.00}$ & $\mathrm{0.00}$ & $\mathrm{0.00}$ & $\mathrm{0.00}$ & $\mathrm{0.00}$ & $\mathrm{0.00}$ \\
& ${\mathrm{DCAC}^{{\text{DD}}_{\text{opt}}}}$
& $\mathbf{0.00}$ & $\mathbf{0.00}$ & $\mathbf{0.00}$ & $\mathbf{0.00}$ & $\mathbf{0.00}$ & $\mathbf{0.00}$ & $\mathbf{0.00}$ & $\mathbf{0.00}$ \\
\midrule

\multirow{3}{*}{\texttt{pegase\_2869}}
& $\mathrm{DCAC}^{\text{BASE}}$
& $\mathrm{0.51}$ & $\mathrm{2.63}$ & $\mathrm{15.69}$ & $\mathrm{13.73}$ & $\mathrm{0.00}$ & $\mathrm{0.00}$ & $\mathrm{0.00}$ & $\mathrm{0.00}$ \\
& ${\mathrm{DCAC}^{{\text{DD}}_{\text{init}}}}$
& $\mathrm{0.00}$ & $\mathrm{0.00}$ & $\mathrm{0.00}$ & $\mathrm{0.00}$ & $\mathrm{0.00}$ & $\mathrm{0.00}$ & $\mathrm{0.00}$ & $\mathrm{0.00}$ \\
& ${\mathrm{DCAC}^{{\text{DD}}_{\text{opt}}}}$
& $\mathbf{0.00}$ & $\mathbf{0.00}$ & $\mathbf{0.00}$ & $\mathbf{0.00}$ & $\mathbf{0.00}$ & $\mathbf{0.00}$ & $\mathbf{0.00}$ & $\mathbf{0.00}$ \\
\midrule

\multirow{3}{*}{\texttt{pegase\_9241}}
& $\mathrm{DCAC}^{\text{BASE}}$
& $\mathrm{2.31}$ & $\mathrm{14.66}$ & $\mathrm{14.74}$ & $\mathrm{10.71}$ & $\mathrm{17.23}$ & $\mathrm{0.07}$ & $\mathrm{2.31}$ & $\mathrm{44.73}$ \\
& ${\mathrm{DCAC}^{{\text{DD}}_{\text{init}}}}$
& $\mathrm{0.00}$ & $\mathrm{0.00}$ & $\mathrm{0.00}$ & $\mathrm{0.00}$ & $\mathrm{0.89}$ & $\mathrm{0.03}$ & $\mathrm{0.00}$ & $\mathrm{0.00}$ \\
& ${\mathrm{DCAC}^{{\text{DD}}_{\text{opt}}}}$
& $\mathbf{0.00}$ & $\mathbf{0.00}$ & $\mathbf{0.00}$ & $\mathbf{0.00}$ & $\mathbf{1.02}$ & $\mathbf{0.04}$ & $\mathbf{0.00}$ & $\mathbf{0.00}$ \\
\bottomrule
\end{tabular}}}
\end{table*}

\begin{table*}[!t]
\centering
\captionsetup{justification=centering,singlelinecheck=false}
\caption{{\small\scshape Performance Table for $\mathrm{DCAC}^{\text{BASE}}$, $\mathrm{DCAC}^{\text{DD}_{\text{init}}}$~\cite{BoatengBentMisraPareekVanHentenryckMolzahn2025TowardsACFeasibility}, and $\mathrm{DCAC}^{\text{DD}_{\text{opt}}}$ across $\mathrm{1{,}000}$ samples ($\mathrm{DC}_{\text{BASE}}$ setpoints)}}
\label{tab:big_eval_dd_perf_dcbase}
\LARGE

\setlength{\tabcolsep}{26pt}
\renewcommand{\arraystretch}{0.8}
\setlength{\extrarowheight}{1.2pt}
\setlength{\aboverulesep}{3pt}
\setlength{\belowrulesep}{2.5pt}

\resizebox{\textwidth}{!}{
\scalebox{1}[0.95]{
\begin{tabular}{llcccccccc}
\toprule
\multirow{2}{*}{\textbf{Test case}} & \multirow{2}{*}{\textbf{Method}} &
\multicolumn{2}{c}{\textbf{Cost Diff. (\%)}} &
\multicolumn{2}{c}{\textbf{MAE$^{\star}$ (p.u.)}} & 
\multicolumn{2}{c}{\textbf{Iter. Count}} &
\multicolumn{2}{c}{\textbf{Solve Time (s)}} \\
\cmidrule(lr){3-4}\cmidrule(lr){5-6}\cmidrule(lr){7-8}\cmidrule(lr){9-10}
& &
val. & \%Improv. &
val. & \%Improv. &
val. & \%Improv. &
val. & \%Improv. \\
\midrule
\midrule

\multirow{3}{*}{\texttt{ieee\_118}}
& $\mathrm{DCAC}^{\text{BASE}}$
& $1.488$ & $\mathbf{+31.0}$ & $0.025$ & $\mathbf{+8.0}$ & $\mathrm{4.0}$ & $\text{---}$ & $0.073$ & $\text{---}$ \\
& ${\mathrm{DCAC}^{{\text{DD}}_{\text{init}}}}$
& $1.035$ & $\mathbf{+0.8}$ & $0.023$ & $0.0$ & $6.2$ & $-1.6$ & $0.319$ & $\mathbf{+1.3}$ \\
& ${\mathrm{DCAC}^{{\text{DD}}_{\text{opt}}}}$
& $1.027$ & $\text{---}$ & $0.023$ & $\text{---}$ & $6.3$ & $\text{---}$ & $0.315$ & $\text{---}$ \\
\midrule

\multirow{3}{*}{\texttt{ACTIVSg\_200}}
& $\mathrm{DCAC}^{\text{BASE}}$
& $0.035$ & $\text{---}$ & $0.003$ & $\mathbf{+33.3}$ & $\mathrm{2.0}$ & $\text{---}$ & $0.065$ & $\text{---}$ \\
& ${\mathrm{DCAC}^{{\text{DD}}_{\text{init}}}}$
& $0.522$ & $-14.2$ & $0.003$ & $\mathbf{+33.3}$ & $4.0$ & $-50.0$ & $0.186$ & $-14.5$ \\
& ${\mathrm{DCAC}^{{\text{DD}}_{\text{opt}}}}$
& $0.596$ & $\text{---}$ & $0.002$ & $\text{---}$ & $6.0$ & $\text{---}$ & $0.213$ & $\text{---}$ \\
\midrule

\multirow{3}{*}{\texttt{ACTIVSg\_500}}
& $\mathrm{DCAC}^{\text{BASE}}$
& $1.383$ & $-27.4$ & $0.010$ & $0.0$ & $\mathrm{3.0}$ & $\text{---}$ & $0.113$ & $\text{---}$ \\
& ${\mathrm{DCAC}^{{\text{DD}}_{\text{init}}}}$
& $0.160$ & $\text{---}$ & $0.010$ & $0.0$ & $4.1$ & $-53.7$ & $0.288$ & $-30.6$ \\
& ${\mathrm{DCAC}^{{\text{DD}}_{\text{opt}}}}$
& $1.762$ & $\text{---}$ & $0.010$ & $\text{---}$ & $6.3$ & $\text{---}$ & $0.376$ & $\text{---}$ \\
\midrule

\multirow{3}{*}{\texttt{pegase\_1354}}
& $\mathrm{DCAC}^{\text{BASE}}$
& $0.394$ & $\mathbf{+23.9}$ & $0.050$ & $\mathbf{+6.0}$ & $\mathrm{4.0}$ & $\text{---}$ & $0.361$ & $\text{---}$ \\
& ${\mathrm{DCAC}^{{\text{DD}}_{\text{init}}}}$
& $0.354$ & $\mathbf{+15.3}$ & $0.049$ & $\mathbf{+4.1}$ & $9.7$ & $\mathbf{+16.5}$ & $1.496$ & $0.0$ \\
& ${\mathrm{DCAC}^{{\text{DD}}_{\text{opt}}}}$
& $0.300$ & $\text{---}$ & $0.047$ & $\text{---}$ & $8.1$ & $\text{---}$ & $1.496$ & $\text{---}$ \\
\midrule

\multirow{3}{*}{\texttt{pegase\_2869}}
& $\mathrm{DCAC}^{\text{BASE}}$
& $0.458$ & $\mathbf{+36.0}$ & $0.042$ & $\mathbf{+9.5}$ & $\mathrm{4.0}$ & $\text{---}$ & $0.915$ & $\text{---}$ \\
& ${\mathrm{DCAC}^{{\text{DD}}_{\text{init}}}}$
& $0.406$ & $\mathbf{+27.8}$ & $0.041$ & $\mathbf{+7.3}$ & $13.6$ & $\mathbf{+25.0}$ & $4.168$ & $\mathbf{+3.3}$ \\
& ${\mathrm{DCAC}^{{\text{DD}}_{\text{opt}}}}$
& $0.293$ & $\text{---}$ & $0.038$ & $\text{---}$ & $10.2$ & $\text{---}$ & $4.031$ & $\text{---}$ \\
\midrule

\multirow{3}{*}{\texttt{pegase\_9241}}
& $\mathrm{DCAC}^{\text{BASE}}$
& $0.631$ & $\mathbf{+17.4}$ & $0.038$ & $\mathbf{+13.2}$ & $\mathrm{3.0}$ & $\text{---}$ & $3.389$ & $\text{---}$ \\
& ${\mathrm{DCAC}^{{\text{DD}}_{\text{init}}}}$
& $0.533$ & $\mathbf{+2.3}$ & $0.036$ & $\mathbf{+8.3}$ & $16.4$ & $\mathbf{+6.7}$ & $15.410$ & $\mathbf{+17.3}$ \\
& ${\mathrm{DCAC}^{{\text{DD}}_{\text{opt}}}}$
& $\mathrm{0.521}$ & $\text{---}$ & $\mathrm{0.033}$ & $\text{---}$ & $15.3$ & $\text{---}$ & $\mathrm{12.743}$ & $\text{---}$ \\
\bottomrule
\end{tabular}}}
\vspace{0.05em}
\begin{flushleft}
\scriptsize
\noindent
The \% Improv. report values relative to $\mathrm{DCAC}^{\text{DD}_{\text{opt}}}$ for each metric. $\mathrm{DCAC}^{\text{DD}_{\text{opt}}}$ uses optimized values \textit{from} $\mathrm{DCAC}^{\text{SS}_{\text{opt}}}$. See Section~\ref{sec:acpf_tuneable_params}. The initialized discrete pipeline, $\mathrm{DCAC}^{\text{DD}_{\text{init}}}$, uses initial parameter values: voltage setpoints initialized to $\mathrm{1.0}$~p.u., with participation factor $\mathrm{1.0}$ assigned to the slack bus and $\mathrm{0.0}$ to all other buses.
\end{flushleft}
\vspace{-0.6em}
\end{table*}

\setlength{\textfloatsep}{6pt}
\setlength{\floatsep}{6pt}
\setlength{\intextsep}{6pt}

\begin{table}[!t]
\centering
\renewcommand{\arraystretch}{1}
\captionsetup{justification=centering, singlelinecheck=false}
\caption{{\small\scshape Solve Time of Optimized $\mathrm{DCAC}$ variants Under Base Case (Using $\mathrm{DC}_{\text{BASE}}$ Setpoints)}}
\label{tab:dcac_optimized_solve_time_only}
\scriptsize
\setlength{\tabcolsep}{6pt}

\begin{tabular}{lccc}
\toprule
\textbf{Test case}
& {\scriptsize ${\mathrm{DCAC}^{{\text{SS}}_{\text{opt}}}}\mathrm{(s)}$}
& {\scriptsize ${\mathrm{DCAC}^{{\text{DD}}_{\text{opt}}}} \mathrm{(s)}$}
& {\scriptsize ${\mathrm{ACOPF}}\mathrm{(s)}$} \\
\midrule
\midrule
\texttt{ieee\_118}     & $\mathbf{0.20}$ & $\mathrm{0.21}$ {\scriptsize $(\mathbf{+4.8})$} & $\mathrm{0.64}$ {\scriptsize $(\mathbf{+68.8})$} \\
\texttt{ACTIVSg\_200}  & $\mathbf{0.16}$ & $\mathrm{0.20}$ {\scriptsize $(\mathbf{+20.0})$} & $\mathrm{0.36}$ {\scriptsize $(\mathbf{+55.6})$} \\
\texttt{ACTIVSg\_500}  & $\mathbf{0.24}$ & $\mathrm{0.32}$ {\scriptsize $(\mathbf{+25.0})$} & $\mathrm{1.48}$ {\scriptsize $(\mathbf{+83.8})$} \\
\texttt{pegase\_1354}  & $\mathbf{0.78}$ & $\mathrm{1.53}$ {\scriptsize $(\mathbf{+49.0})$} & $\mathrm{2.79}$ {\scriptsize $(\mathbf{+72.0})$} \\
\texttt{pegase\_2869}  & $\mathbf{1.92}$ & $\mathrm{3.76}$ {\scriptsize $(\mathbf{+48.9})$} & $\mathrm{4.93}$ {\scriptsize $(\mathbf{+61.1})$} \\
\texttt{pegase\_9241}  & $\mathbf{7.62}$ & $\mathrm{11.97}$ {\scriptsize $(\mathbf{+36.3})$} & $\mathrm{30.28}$ {\scriptsize $(\mathbf{+74.8})$} \\
\bottomrule
\end{tabular}
\begin{flushleft}
\vspace{-0.2em}
\scriptsize{
Parentheses show $\mathrm{DCAC}^{\text{SS}_{\text{opt}}}$ solve-time \% Improv. relative to that method. $\mathrm{DCAC}^{\text{SS}_{\text{opt}}}$ uses the smooth PACR approach, while $\mathrm{DCAC}^{\text{DD}_{\text{opt}}}$ uses discrete mode.
}
\end{flushleft}
\vspace{-0.5em}
\end{table}

\vspace{-1em}

\subsection{AC Feasibility Restoration Assessment: Initialized versus Optimized Smooth Pipelines (($\mathrm{DCAC}^{\mathrm{SS}_\text{init}}$) vs. ($\mathrm{DCAC}^{\mathrm{SS}_\text{opt}}$)):}
\label{sec:ac_ss}

This subsection evaluates the effect of parameter optimization within the smooth AC restoration pipeline. Specifically, the initialized smooth pipeline, $\mathrm{DCAC}^{\mathrm{SS}_{\text{init}}}$ is compared against the optimized smooth pipeline, $\mathrm{DCAC}^{\mathrm{SS}_{\text{opt}}}$, with $\mathrm{DCAC}^{\mathrm{BASE}}$ included as a conventional single-slack reference. The goal is to assess whether the learned smooth distributed-slack and PV/PQ regulation parameters improve AC feasibility restoration from the lossless initialized pipeline, while maintaining favorable cost, dispatch-error, and computational performance.

\textit{1) Constraint Violation Table Analysis:}
Table~\ref{tab:big_eval_ss_viol_dcbase} compares active-power, reactive-power, voltage-limit, and line-limit violations for $\mathrm{DCAC}^{\text{BASE}}$, $\mathrm{DCAC}^{\mathrm{SS}_{\text{init}}}$, and $\mathrm{DCAC}^{\mathrm{SS}_{\text{opt}}}$ across $\mathrm{1{,}000}$ test samples. The single-slack baseline frequently leaves residual AC infeasibilities, especially reactive-power violations,with average violation rates of $\mathrm{18.46\%}$, $\mathrm{15.69\%}$, and $\mathrm{14.74\%}$ on \texttt{pegase\_1354}, \texttt{pegase\_2869}, and \texttt{pegase\_9241}, respectively. In contrast, both smooth pipelines eliminate active- and reactive-power violations across all tested systems. The optimized smooth pipeline further improves the remaining feasibility behavior relative to the initialized smooth pipeline. It removes the small line-limit violations observed for $\mathrm{DCAC}^{\mathrm{SS}_{\text{init}}}$ on \texttt{ACTIVSg\_500} and reduces the average voltage-violation rates from $\mathrm{0.21\%}$ to $\mathrm{0.03\%}$ on \texttt{pegase\_2869} and from $\mathrm{3.75\%}$ to $\mathrm{2.61\%}$ on \texttt{pegase\_9241}. Thus, parameter optimization preserves major feasibility gains of the initialized smooth pipeline while further reducing residual voltage and line-limit violations in the more challenging cases.

\textit{2) Evaluation-Loss and Performance Table Analysis:}
Table~\ref{tab:dcac_ss_loss} reports the evaluation loss~(\ref{eq:loss_sl}) for the baseline and smooth pipelines. Compared with $\mathrm{DCAC}^{\mathrm{SS}_{\text{init}}}$, the optimized smooth pipeline reduces the evaluation loss on most systems, including \texttt{ACTIVSg\_200}, \texttt{ACTIVSg\_500}, \texttt{pegase\_2869}, and \texttt{pegase\_9241}. For example, the loss decreases from $\mathrm{0.042}$ to $\mathrm{0.025}$ on \texttt{pegase\_9241} and from $\mathrm{0.041}$ to $\mathrm{0.031}$ on \texttt{pegase\_2869}. The baseline produces much larger losses across all cases, confirming that the smooth restoration structure substantially improves recovery quality over conventional single-slack ACPF. This is consistent with the findings in \cite{BoatengBentMisraPareekVanHentenryckMolzahn2025TowardsACFeasibility}. Table~\ref{tab:big_eval_ss_perf_dcbase} further shows that optimization improves cost consistency, dispatch accuracy, and computational performance on the larger systems. On \texttt{pegase\_1354}, $\mathrm{DCAC}^{\mathrm{SS}_{\text{opt}}}$ reduces the CD from $\mathrm{0.333\%}$ to $\mathrm{0.044\%}$ relative to $\mathrm{DCAC}^{\mathrm{SS}_{\text{init}}}$, while reducing MAE from $\mathrm{0.048}$ to $\mathrm{0.040}$~p.u. On \texttt{pegase\_2869}, the cost difference decreases from $\mathrm{0.387\%}$ to $\mathrm{0.021\%}$ and MAE reduces from $\mathrm{0.041}$ to $\mathrm{0.030}$~p.u. On \texttt{pegase\_9241}, the cost difference is lowered also, from $\mathrm{0.510\%}$ to $\mathrm{0.124\%}$ and MAE follows that trend, from $\mathrm{0.035}$ to $\mathrm{0.025}$~p.u. The optimized smooth pipeline also lowered average solve time on these large cases. These results show that optimizing the smooth restoration parameters improves both AC feasibility restoration and ACOPF-tracking performance relative to the initialized smooth pipeline, while remaining substantially better than the single-slack baseline.

\subsection{AC Feasibility Restoration Assessment: Initialized versus Optimized Discrete Pipelines (($\mathrm{DCAC}^{\mathrm{DD}_\text{init}}$) vs. ($\mathrm{DCAC}^{\mathrm{DD}_\text{opt}}$)):}
\label{sec:ac_ss2}

This subsection evaluates the effect of transferring the optimized parameters learned from the smooth PACR formulation to the corresponding discrete restoration pipeline. Specifically, the initialized discrete pipeline, $\mathrm{DCAC}^{\mathrm{DD}_{\text{init}}}$~\cite{BoatengBentMisraPareekVanHentenryckMolzahn2025TowardsACFeasibility}, is compared against the optimized discrete pipeline, $\mathrm{DCAC}^{\mathrm{DD}_{\text{opt}}}$, with $\mathrm{DCAC}^{\mathrm{BASE}}$ included as a conventional single-slack reference. This was done to assess whether the learned distributed-slack participation factors and voltage-setpoint values improve AC feasibility restoration in the discrete ACPF procedure relative to the initialized discrete pipeline.

\begin{figure*}[!t]
  \centering
  \begin{subfigure}{0.49\textwidth}
    \centering
    \includegraphics[width=\linewidth]{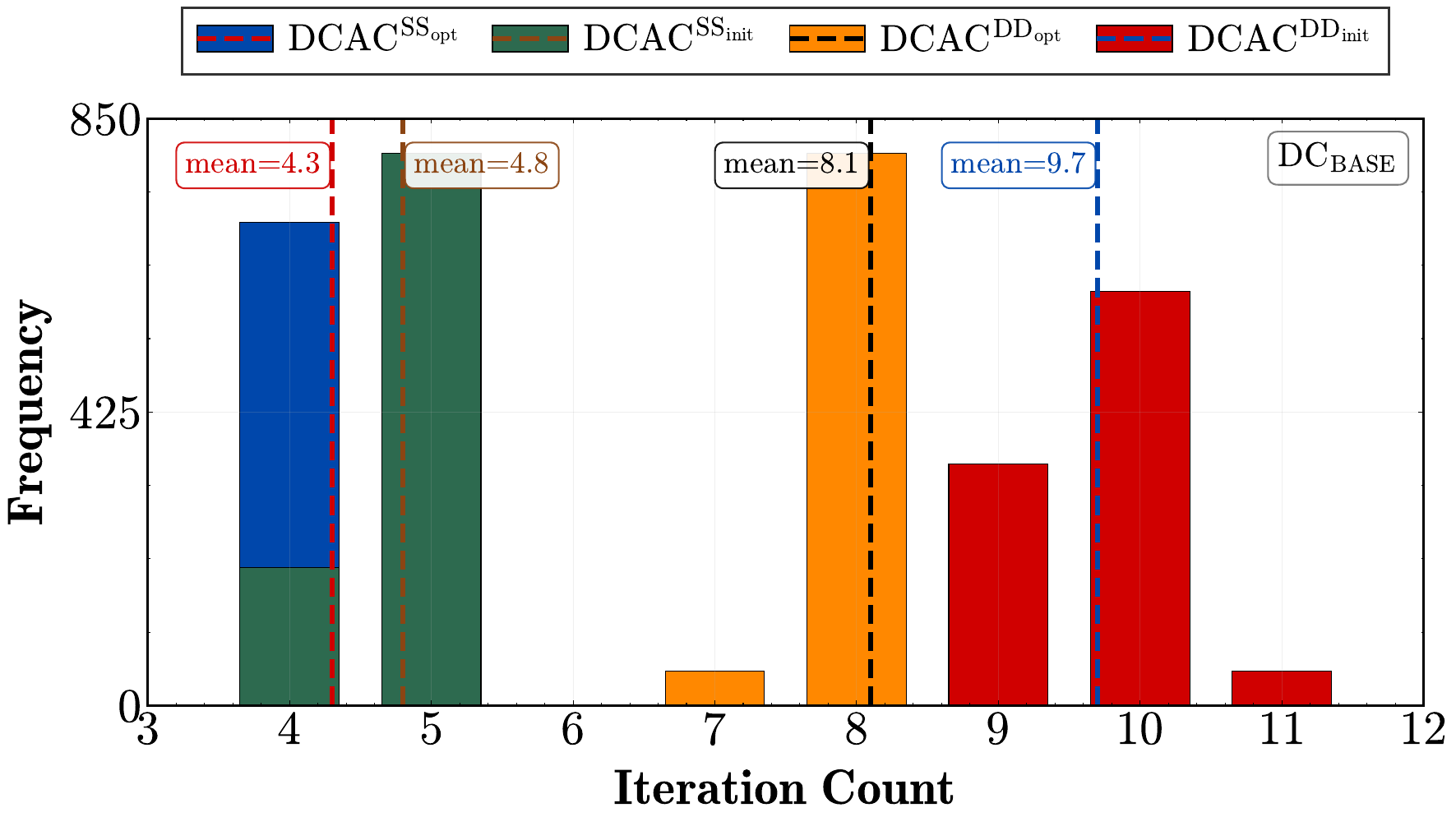}
    \caption{\textbf{Iteration count for \texttt{pegase\_1354} on $\mathbf{1{,}000}$ samples.}}
    \label{fig:iter-1354}
  \end{subfigure}\hfill
  \begin{subfigure}{0.49\textwidth}
    \centering
    \includegraphics[width=\linewidth]{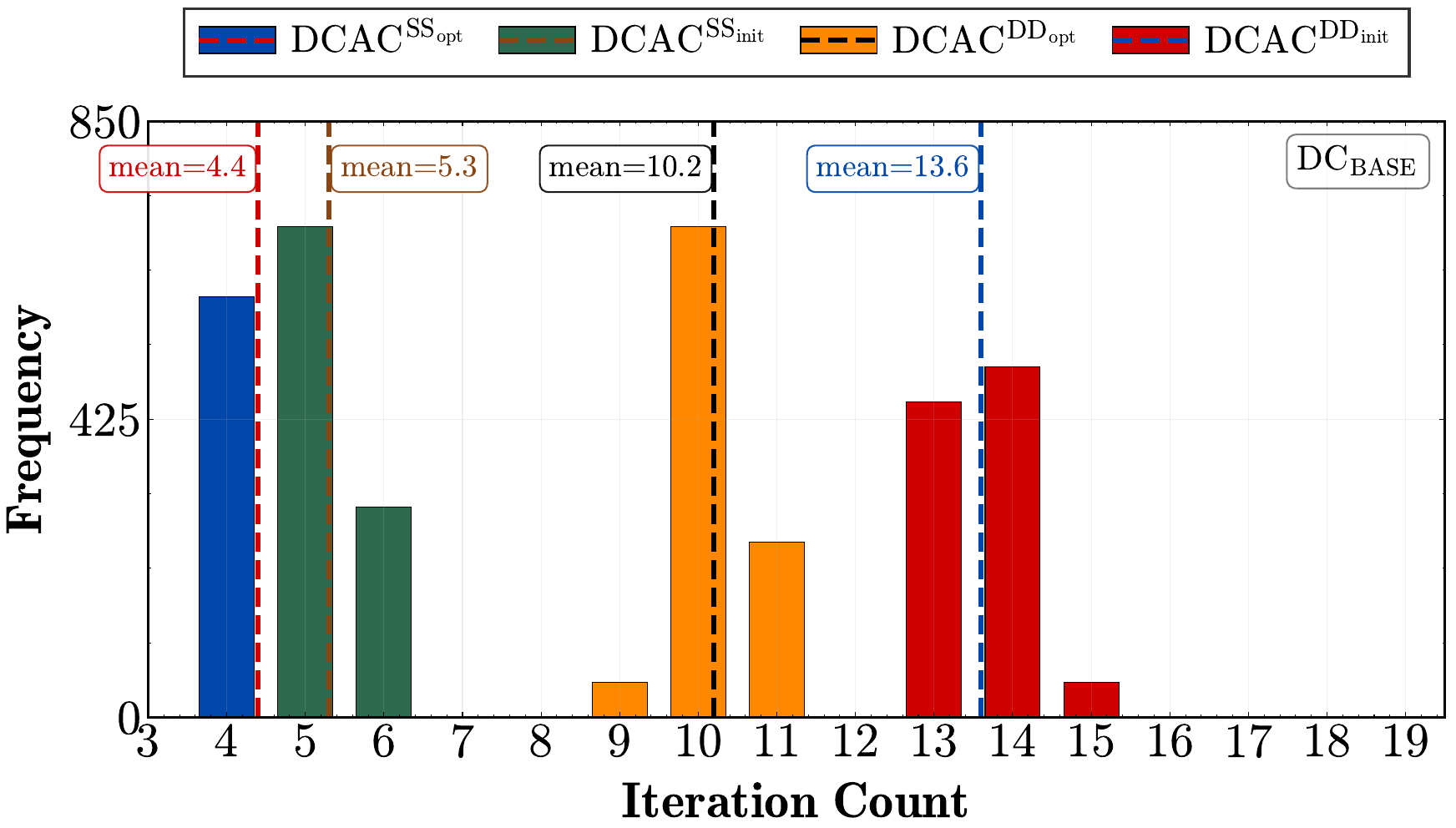}
    \caption{\textbf{Iteration count for \texttt{pegase\_2869} on $\mathbf{1{,}000}$ samples.}}
    \label{fig:iter-2869}
  \end{subfigure}

  \vspace{-0.2em}
  \caption{Iteration count comparison for the four $\mathrm{DCAC}$ variants under load uncertainty using $\mathrm{1{,}000}$ samples per case.
  (a) \texttt{pegase\_1354}. 
  (b) \texttt{pegase\_2869}. 
  The compared methods are
    $\mathrm{DCAC}^{{\mathrm{SS}}_{\text{opt}}}$
  (\textbf{\textcolor{blue}{blue}}),
  $\mathrm{DCAC}^{{\mathrm{SS}}_{\text{init}}}$
  (\textbf{\textcolor{ACgreen}{green}}),
  $\mathrm{DCAC}^{{\mathrm{DD}}_{\text{opt}}}$
  (\textbf{\textcolor{orange}{orange}}), and
  $\mathrm{DCAC}^{{\mathrm{DD}}_{\text{init}}}$~\cite{BoatengBentMisraPareekVanHentenryckMolzahn2025TowardsACFeasibility}
  (\textbf{\textcolor{red}{red}}). The \textit{`opt'} variant uses parameters trained via Algorithm~\ref{alg:acpf_train}, and `init' uses initial flat-start parameter values.} 
  \label{fig:iteration-count-comparison}
  \vspace{-0.4em}
\end{figure*}

\begin{figure*}[!t]
  \centering
  \begin{subfigure}{0.49\textwidth}
    \centering
    \includegraphics[width=\linewidth]{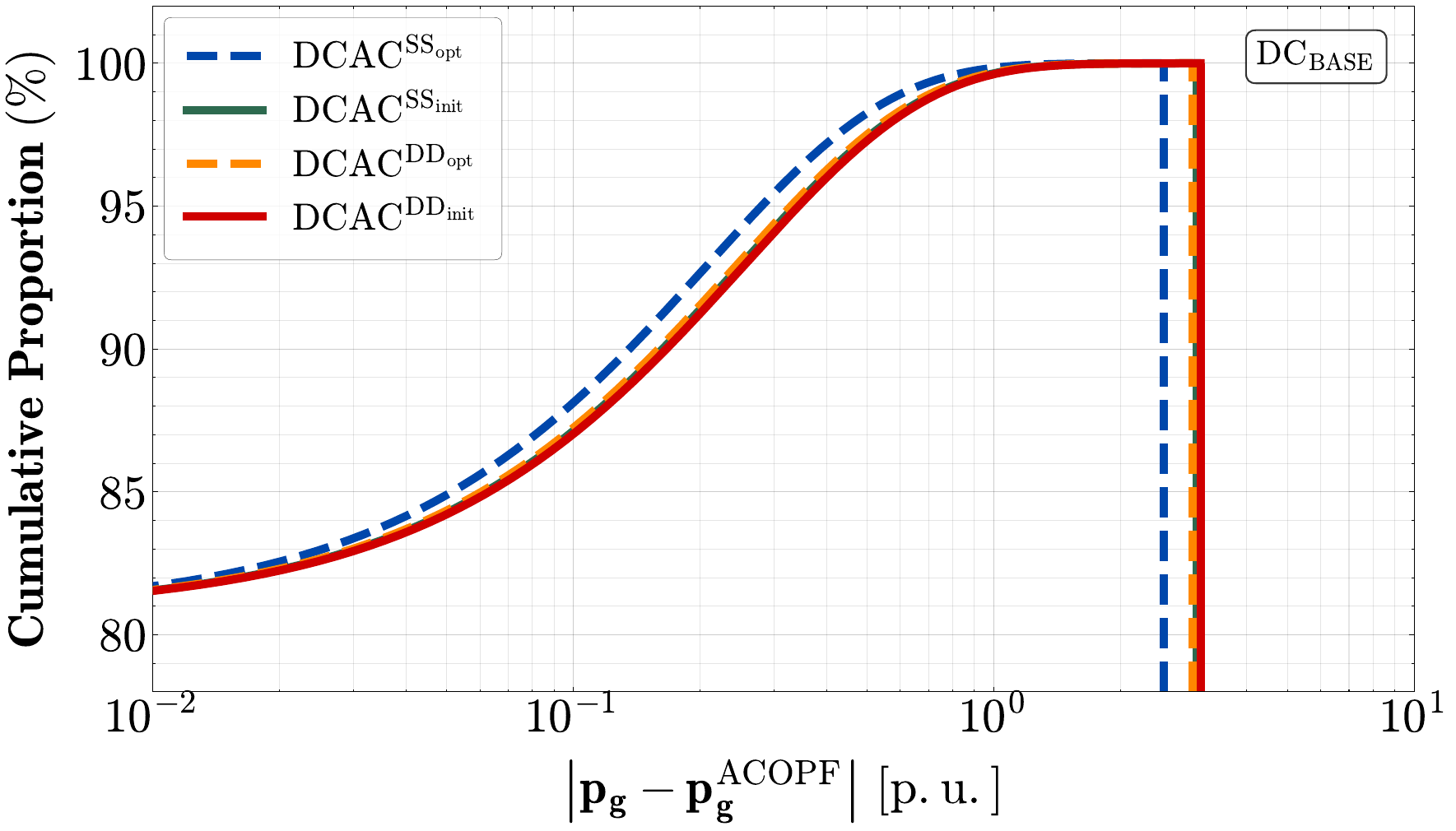}
    \caption{\textbf{Cumulative plot of $\mathbf{p}_{\mathbf{g}}$ error for \texttt{pegase\_2869} on $\mathbf{1{,}000}$ samples.}}
    \label{fig:cumulative_plot_2869_pg_and_vma}
  \end{subfigure}\hfill
  \begin{subfigure}{0.49\textwidth}
    \centering
    \includegraphics[width=\linewidth]{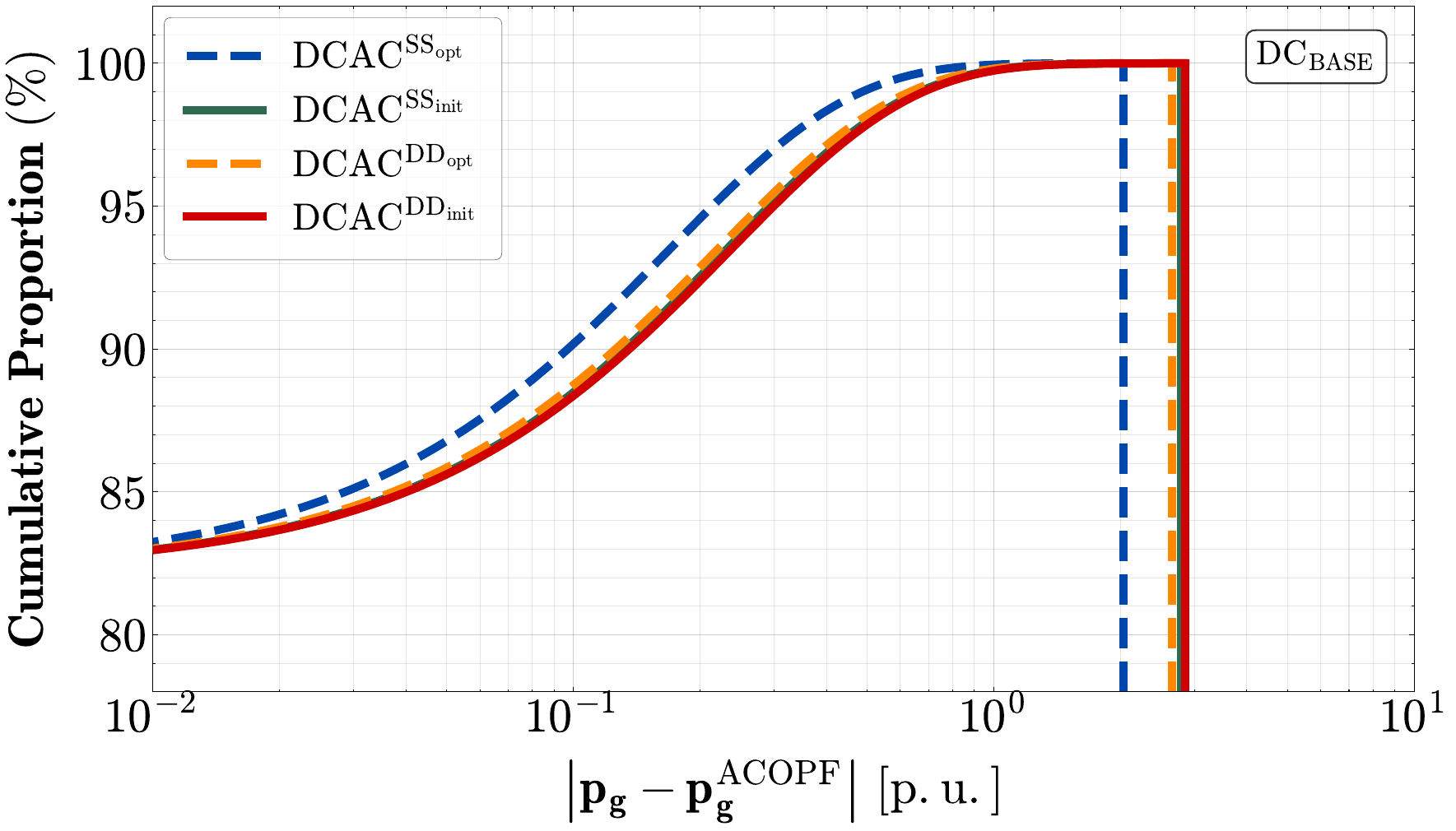}
    \caption{\textbf{Cumulative plot of $\mathbf{p}_{\mathbf{g}}$ error for \texttt{pegase\_9241} on $\mathbf{1{,}000}$ samples.}}
    \label{fig:cumulative_plot_2869_pg_and_vmb}
  \end{subfigure}
  \vspace{-0.2em}
  \caption{Empirical cumulative distribution of active-power dispatch error,
  $|\mathbf{p}_{\mathbf{g}}-\mathbf{p}_{\mathbf{g}}^{\mathrm{ACOPF}}|$, using $\mathrm{1{,}000}$ samples per case.
  (a) \texttt{pegase\_2869}. 
  (b) \texttt{pegase\_9241}. 
  The DCAC methods are
  $\mathrm{DCAC}^{{\mathrm{SS}}_{\text{opt}}}$
  (\textbf{\textcolor{blue}{blue}}),
  $\mathrm{DCAC}^{{\mathrm{SS}}_{\text{init}}}$
  (\textbf{\textcolor{ACgreen}{green}}),
  $\mathrm{DCAC}^{{\mathrm{DD}}_{\text{opt}}}$
  (\textbf{\textcolor{orange}{orange}}), and
  $\mathrm{DCAC}^{{\mathrm{DD}}_{\text{init}}}$~\cite{BoatengBentMisraPareekVanHentenryckMolzahn2025TowardsACFeasibility}
  (\textbf{\textcolor{red}{red}}). The \textit{`opt'} variant uses parameters trained via Algorithm~\ref{alg:acpf_train}, and `init' uses initial flat-start parameter values.}
  \label{fig:ecdf-errors-2869}
  \vspace{-0.5em}
\end{figure*}

\textit{1) Constraint Violation Table Analysis:}
Table~\ref{tab:big_eval_dd_viol_dcbase} reports constraint violations for $\mathrm{DCAC}^{\text{BASE}}$, $\mathrm{DCAC}^{\text{DD}_{\text{init}}}$, and $\mathrm{DCAC}^{\text{DD}_{\text{opt}}}$. As in the smooth-pipeline comparison in Section~\ref{sec:ac_ss}, the single-slack baseline leaves significant residual infeasibilities, including reactive-power violation rates of $\mathrm{8.03\%}$ on \texttt{ACTIVSg\_500} and $\mathrm{18.46\%}$ on \texttt{pegase\_1354}. On the largest case, \texttt{pegase\_9241}, the baseline also produces average active-power, voltage-limit, and line-limit violations.

Both discrete restoration pipelines substantially improve feasibility relative to this baseline. The initialized discrete pipeline already removes all active- and reactive-power violations, but small residual voltage or line-limit violations remain on a few cases. The optimized discrete pipeline improves several of these residual violations: it removes the small voltage violation on \texttt{ACTIVSg\_200}, removes the line-limit violation on \texttt{ACTIVSg\_500}, and achieves zero violations on \texttt{ieee\_118}, \texttt{ACTIVSg\_200}, \texttt{ACTIVSg\_500}, \texttt{pegase\_1354}, and \texttt{pegase\_2869}. On \texttt{pegase\_9241}, it eliminates active-power, reactive-power, and line-limit violations, while leaving a small average voltage-violation rate of $\mathrm{1.02\%}$. These results show that the optimized parameters learned in the smooth formulation transfer effectively to the discrete restoration setting. This is important because many existing ACPF tools use discrete PV/PQ switching and distributed-slack logic; therefore, the learned parameters can improve standard discrete recovery procedures without requiring a fully differentiable solver implementation.

\textit{2) Performance Table Analysis:}
Table~\ref{tab:big_eval_dd_perf_dcbase}  compares cost difference, generator active-power MAE, iteration count, and solve time for the baseline and discrete restoration pipelines. The optimized discrete pipeline improves the restored operating point relative to the initialized discrete pipeline on the larger PEGASE systems. On \texttt{pegase\_1354}, the cost difference decreases from $\mathrm{0.354\%}$ to $\mathrm{0.300\%}$ and MAE decreases from $\mathrm{0.049}$ to $\mathrm{0.047}$~p.u. On \texttt{pegase\_9241}, MAE decreases from $\mathrm{0.036}$ to $\mathrm{0.033}$~p.u., while solve time is lowered from $\mathrm{15.410}$~s to $\mathrm{12.743}$~s. Relative to the single-slack baseline, $\mathrm{DCAC}^{\mathrm{DD}_{\text{opt}}}$ provides stronger feasibility restoration and better dispatch accuracy on the large cases. For \texttt{pegase\_2869}, it attains a $\mathrm{0.038}$~p.u. MAE and $\mathrm{0.293\%}$ cost difference, compared with $\mathrm{0.042}$~p.u. and $\mathrm{0.458\%}$ for $\mathrm{DCAC}^{\text{BASE}}$. For \texttt{pegase\_9241}, it achieves $\mathrm{0.033}$~p.u. MAE and $\mathrm{0.521\%}$ cost difference, compared with $\mathrm{0.038}$~p.u. and $\mathrm{0.631\%}$ for the baseline. Although the discrete restoration pipeline requires more Newton iterations than the single-slack ACPF baseline, it produces a much stronger AC feasibility recovery, confirming that optimized distributed slack and voltage regulation improve the quality of the restored AC operating point.

\vspace{-0.55em}

\subsection{Complete Performance Comparison of Discrete and Smooth DCAC Pipelines (($\mathrm{DCAC}^{\mathrm{DD}_\text{}}$)  vs. ($\mathrm{DCAC}^{\mathrm{SS}_\text{}}$)):}
\label{sec:complete_perf}

This subsection directly compares the proposed smooth restoration pipeline against the discrete restoration pipeline~\cite{BoatengBentMisraPareekVanHentenryckMolzahn2025TowardsACFeasibility}, after parameter optimization. The purpose is to separate the benefit of the learned parameters from the additional benefit of using a smooth differentiable ACPF formulation. The results show that although both optimized pipelines improve restoration quality, as shown in Sections \ref{sec:ac_ss} and \ref{sec:ac_ss2}, the smooth pipeline achieves better accuracy, with substantially lower computational effort than our former discrete framework ~\cite{BoatengBentMisraPareekVanHentenryckMolzahn2025TowardsACFeasibility}.

\textit{1) Optimized Pipeline Solve-Time Analysis:}
Table~\ref{tab:dcac_optimized_solve_time_only} compares the solve times of $\mathrm{DCAC}^{\text{SS}_{\text{opt}}}$, $\mathrm{DCAC}^{\text{DD}_{\text{opt}}}$, and ACOPF across all test cases using $\mathrm{DC}_{\text{BASE}}$ setpoints. The optimized smooth pipeline is the fastest method in every case, with solve times ranging from $\mathrm{0.16}$~s on \texttt{ACTIVSg\_200} to $\mathrm{7.62}$~s on \texttt{pegase\_9241}. Relative to the optimized discrete pipeline, $\mathrm{DCAC}^{\text{SS}_{\text{opt}}}$ reduces solve time by $\mathrm{4.8\%}$--$\mathrm{49.0\%}$. Relative to ACOPF, the reduction is larger, ranging from $\mathrm{55.6\%}$ to $\mathrm{83.8\%}$.

This shows that the smooth formulation improves computational efficiency while preserving the speed advantage of DC-based dispatch followed by AC feasibility restoration. The gains over $\mathrm{DCAC}^{\text{DD}_{\text{opt}}}$ suggest that avoiding discrete distributed-slack and PV/PQ switching updates reduces the computational burden of the restoration step.

\textit{2) Iteration Count Figure Analysis:}
Figures~\ref{fig:iter-1354} and~\ref{fig:iter-2869} compare Newton iteration-count distributions for \texttt{pegase\_1354} and \texttt{pegase\_2869} across $\mathrm{1{,}000}$ load-uncertainty scenarios using $\mathrm{DC}_{\text{BASE}}$ setpoints. The optimized smooth pipeline $\mathrm{DCAC}^{{\text{SS}}_{\text{opt}}}$ concentrates most of its probability mass at the lowest iteration counts, with mean values of approximately $\mathrm{4.3}$ and $\mathrm{4.4}$ iterations on \texttt{pegase\_1354} and \texttt{pegase\_2869}, respectively. The initialized smooth version $\mathrm{DCAC}^{{\text{SS}}_{\text{init}}}$ remains competitive, with means of $\mathrm{4.8}$ and $\mathrm{5.3}$ iterations, indicating that the smooth formulation itself provides a favorable numerical structure even before training. By contrast, the discrete pipelines require substantially more iterations. On \texttt{pegase\_1354}, $\mathrm{DCAC}^{{\text{DD}}_{\text{opt}}}$ and $\mathrm{DCAC}^{{\text{DD}}_{\text{init}}}$ average approximately $\mathrm{8.1}$ and $\mathrm{9.7}$ iterations, respectively. On \texttt{pegase\_2869}, the corresponding averages are approximately $\mathrm{10.2}$ and $\mathrm{13.6}$. 

These results highlight a tradeoff between the two implementations. The smooth pipeline provides the best accuracy and speed because it avoids outer-loop bus-type switching and discontinuous PV/PQ updates~\cite{zeng2023pvpq,zhao2008bus}. However, the discrete pipeline is often more directly compatible with traditional power-flow packages that implement PV/PQ switching and distributed slack through discrete logic. The relatively small degradation of $\mathrm{DCAC}^{{\text{DD}}_{\text{opt}}}$ compared with $\mathrm{DCAC}^{{\text{SS}}_{\text{opt}}}$ therefore suggests that parameters optimized in the smooth formulation can still be useful in standard discrete ACPF implementations.

\textit{3) Cumulative Error Plot Analysis:}
Figures~\ref{fig:cumulative_plot_2869_pg_and_vma} and~\ref{fig:cumulative_plot_2869_pg_and_vmb} report the empirical cumulative distributions of the generator active-power dispatch error,
$\lvert \mathbf{p}_{\mathrm{g}}-\mathbf{p}_{\mathrm{g}}^{\mathrm{ACOPF}}\rvert$, for \texttt{pegase\_2869} and \texttt{pegase\_9241}. Near the $10^{-2}$~p.u. threshold, all variants show similar cumulative behavior. For \texttt{pegase\_2869}, $\mathrm{DCAC}^{{\text{SS}}_{\text{opt}}}$ places approximately $\mathrm{81.9\%}$ of generator errors below this threshold, compared with $\mathrm{81.8\%}$ for $\mathrm{DCAC}^{{\text{DD}}_{\text{opt}}}$ and about $\mathrm{81.6\%}$ for the initialization-based variants. For \texttt{pegase\_9241}, the optimized smooth pipeline again gives the highest proportion, approximately $\mathrm{83.3\%}$, while the other methods remain close at about $\mathrm{83.0\%}$.

The main separation appears in the right tail. On \texttt{pegase\_2869}, the worst-case error for $\mathrm{DCAC}^{{\text{SS}}_{\text{opt}}}$ is approximately $\mathrm{2.53}$~p.u., compared with roughly $\mathrm{2.97}$--$\mathrm{3.17}$~p.u. for the other variants. On \texttt{pegase\_9241}, the optimized smooth pipeline reaches a worst-case error of approximately $\mathrm{2.29}$~p.u., while the other variants lie between about $\mathrm{2.99}$ and $\mathrm{3.22}$~p.u. Thus, the optimized smooth formulation does not only improve average behavior; it also reduces the largest dispatch deviations, which is important for robust restoration under load uncertainty.

\section{Conclusion}
\label{sec:conclusion}


This work develops a smooth, differentiable ACPF restoration learning method (PACR) for mapping DCOPF dispatches to AC-consistent operating points. The method replaces discrete restoration logic in our previous work \cite{BoatengBentMisraPareekVanHentenryckMolzahn2025TowardsACFeasibility}, with smooth surrogates, including softmax-based distributed slack and sigmoid PV/PQ regulation, allowing voltage setpoints and slack-participation parameters to be optimized. Experiments on IEEE, ACTIVSg, and PEGASE systems show that the optimized pipelines improve AC feasibility, reduce Newton iterations, and lower dispatch errors relative to conventional single-slack ACPF recovery. The gains are strongest on large PEGASE cases: the optimized smooth DCAC method reduces cost difference by approximately $\mathrm{95\%}$ on the $\mathrm{2{,}869}$-bus case and $\mathrm{80\%}$ on the $\mathrm{9{,}241}$-bus case, while solving the $\mathrm{9{,}241}$-bus case about $\mathrm{75\%}$ faster than ACOPF. Overall, the optimized smooth pipeline is the best-performing restoration method, and its learned participation factors and voltage setpoints also transfer effectively to the discrete DCAC pipeline \cite{BoatengBentMisraPareekVanHentenryckMolzahn2025TowardsACFeasibility}. 

Future work will extend the differentiable restoration method to additional corrective controls, including transformer taps and capacitor shunts, and will explore end-to-end self-supervised training of the full DCOPF$\rightarrow$ACPF pipeline. This direction connects naturally with recent work on parameterized and optimized DCOPF models~\cite{taheri2024optimizing,rosemberg_klamkin_2025,constanteflores2024acnetwork}, while posing a more demanding learning problem: the DCOPF parameters must be optimized not only for dispatch cost, but also for their downstream effect on AC inequality constraint violations.

\bibliographystyle{ieeetr}
\bibliography{references}


\appendices

\section{Q--V Sigmoid Curve at Voltage Setpoint}
\label{sec:appendix_a}

This appendix verifies that the smooth Q--V regulation law passes through the prescribed setpoint $\bigl(v_i^{sp},\,q_{g,i}^{sp}\bigr)$.

\paragraph*{Claim}
If $v_i=v_i^{sp}$, then $q_{g,i}^{ac}=q_{g,i}^{sp}$.

\paragraph*{Proof}
Let $q_{g,i}^{\min}<q_{g,i}^{sp}<q_{g,i}^{\max}$ and $\psi_s>0$. 
\begin{equation}
\label{eq:qv_sigmoid}
q_{g,i}^{ac}
=
q_{g,i}^{\min}
+
\frac{q_{g,i}^{\max}-q_{g,i}^{\min}}
{
1+\exp\!\left(
\psi_s(v_i-v_i^{sp})
+
\ln\!\frac{q_{g,i}^{\max}-q_{g,i}^{sp}}
{q_{g,i}^{sp}-q_{g,i}^{\min}}
\right)
}.
\end{equation}
At $v_i=v_i^{sp}$,
\begin{equation}
\small
q_{g,i}^{ac}
= q_{g,i}^{\min}
+\frac{q_{g,i}^{\max}-q_{g,i}^{\min}}
{1+\frac{q_{g,i}^{\max}-q_{g,i}^{sp}}{q_{g,i}^{sp}-q_{g,i}^{\min}}}
= q_{g,i}^{\min}+(q_{g,i}^{sp}-q_{g,i}^{\min})
= q_{g,i}^{sp}.
\end{equation}
Hence, the parameterized sigmoid passes through $\bigl(v_i^{sp},\,q_{g,i}^{sp}\bigr)$.

\section{Differentiating Through the Augmented ACPF}
\label{sec:appendix_b}

This appendix provides the mathematical framework for computing gradients through the augmented ACPF system to enable parameter optimization via backpropagation.

\vspace{0.5em}

Let the converged augmented ACPF solution be
$\mathbf{x}_{\mathrm{ac}}=(\boldsymbol{\theta},\mathbf{v},\ell,\mathbf{p}_{\mathrm{g}}^{\mathrm{ac}},\mathbf{q}_{\mathrm{g}}^{\mathrm{ac}})$
and define the trainable parameters
\begin{equation}
\boldsymbol{\xi}\triangleq\{\pi_s,\phi_s,\psi_s,\mathbf{v}^{\mathrm{sp}}\},
\end{equation}
where $\psi_t$ is treated as a fixed (or bounded) parameter that shifts effective reactive limits (see Model~\ref{mod:acpf}).
For each scenario, $\mathbf{x}_{\mathrm{ac}}$ satisfies the implicit system
\begin{equation}
\mathbf{g}(\mathbf{x}_{\mathrm{ac}};\boldsymbol{\xi})
=
\begin{bmatrix}
\Delta \mathbf{p}(\mathbf{x}_{\mathrm{ac}};\boldsymbol{\xi})\\
\Delta \mathbf{q}(\mathbf{x}_{\mathrm{ac}};\boldsymbol{\xi})\\
\ell^{\mathrm{tot}}(\mathbf{x}_{\mathrm{ac}})
\end{bmatrix}
=\mathbf{0}.
\end{equation}
Given the full-batch objective:
\begin{equation}
\mathcal{J}
=
\frac{1}{|\mathcal{D}|}
\sum_{\mathcal{D}}
\left[
w_P
\big\|\mathbf{p}_{\mathrm{g}}^{\mathrm{ac}}-\mathbf{p}_{\mathrm{g}}^{\mathrm{ref}}\big\|_1
+
w_V
\big\|\mathbf{v}-\mathbf{v}^{\mathrm{ref}}\big\|_1
\right]
+
\mathcal{R}(\boldsymbol{\xi}),
\end{equation}
gradients are computed by implicit differentiation. Let
$\mathbf{J}_{\mathrm{ac}}=\partial\mathbf{g}/\partial\mathbf{x}_{\mathrm{ac}}$. For each scenario,
\begin{equation}
\mathbf{J}_{\mathrm{ac}}^{\top}\boldsymbol{\lambda}
=
\left(\frac{\partial \mathcal{L}^{\mathrm{sc}}}{\partial \mathbf{x}_{\mathrm{ac}}}\right)^{\top},
\qquad
\frac{\partial \mathcal{J}}{\partial \boldsymbol{\xi}}
=
-\frac{1}{|\mathcal{D}|}
\sum_{\mathcal{D}}
\boldsymbol{\lambda}^{\top}
\frac{\partial \mathbf{g}}{\partial \boldsymbol{\xi}}
+
\frac{\partial \mathcal{R}}{\partial \boldsymbol{\xi}} .
\label{eq:adjoint_short}
\end{equation}

\vspace{-1em}

\subsection{Where Parameters Enter the Mismatches}

Parameters affect $\Delta\mathbf{p}$ through smooth distributed slack
$p_{\mathrm{g},i}^{\mathrm{dc}}+\alpha_i \ell$ and affect $\Delta\mathbf{q}$ through the smooth Q--V map
$q_{\mathrm{g},i}^{\mathrm{ac}}(v_i)$:
\[
\frac{\partial \Delta p_i}{\partial \alpha_i}=\ell,
\qquad
\frac{\partial \Delta q_i}{\partial \xi}=\frac{\partial q_{\mathrm{g},i}^{\mathrm{ac}}}{\partial \xi}.
\]

\subsection{Differentiable Building Blocks}

\vspace{0.5em}

\paragraph{Softplus headroom}
Let $r_i=p_{\mathrm{g},i}^{\max}-p_{\mathrm{g},i}^{\mathrm{dc}}$ and
$\tilde h_i=\frac{1}{\pi_s}\ln(1+e^{\pi_s r_i})-\frac{\ln 2}{\pi_s}$.
Then
\begin{equation}
\frac{\partial \tilde h_i}{\partial r_i}=\sigma(\pi_s r_i),
\qquad
\sigma(u)=\frac{1}{1+e^{-u}}.
\end{equation}

\vspace{0.5em}

\paragraph{Softmax participation.}
$\alpha_i=\frac{e^{\phi_s\tilde h_i}}{\sum_{j\in\mathcal{G}}e^{\phi_s\tilde h_j}}$ implies
\begin{equation}
\frac{\partial \alpha_i}{\partial \phi_s}
=
\alpha_i\Big(\tilde h_i-\sum_{j\in\mathcal{G}}\alpha_j\tilde h_j\Big),
\qquad
\frac{\partial \alpha_i}{\partial \tilde h_k}
=
\phi_s\,\alpha_i(\mathbb{I}[i=k]-\alpha_k).
\end{equation}
(Chain rule gives $\partial \alpha_i/\partial \pi_s$ through $\tilde h$ if $\pi_s$ is trained.)

\vspace{0.5em}

\paragraph{Sigmoid Q--V regulation}
Let $q_{\mathrm{g},i}^{\mathrm{ac}}=q_{\min,i}^{\mathrm{eff}}+\Delta q_i^{\mathrm{eff}}\,s_i$ where
$s_i=\bigl(1+e^{\zeta_i}\bigr)^{-1}$ and $\zeta_i=\psi_s(v_i-v_i^{\mathrm{sp}})+(\cdots)$.
The $(\cdots)$ term collects optional offsets (e.g., the log-offset enforcing passage through $(v_i^{\mathrm{sp}},q_{\mathrm{g},i}^{\mathrm{sp}})$) and is omitted here for brevity; including it preserves differentiability and only adds chain-rule terms (Model~\ref{mod:acpf}).
Then $s_i(1-s_i)$ provides compact derivatives:
\begin{equation}
\frac{\partial q_{\mathrm{g},i}^{\mathrm{ac}}}{\partial \psi_s}
=
-\Delta q_i^{\mathrm{eff}}\,s_i(1-s_i)(v_i-v_i^{\mathrm{sp}}),
\\
\frac{\partial q_{\mathrm{g},i}^{\mathrm{ac}}}{\partial v_i^{\mathrm{sp}}}
=
\Delta q_i^{\mathrm{eff}}\,\psi_s\,s_i(1-s_i).
\end{equation}

\end{document}